\documentclass[12pt]{article}

\usepackage[T1]{fontenc} 				

\usepackage{amssymb,amsmath,amscd}
\usepackage[english]{babel}
\usepackage{titlesec}
\usepackage{slashed}
\usepackage{hyperref}
\usepackage{graphicx,xcolor}
\usepackage{enumerate}
\usepackage{bbm} 						

\hypersetup{
	colorlinks=true,
	linkcolor=black,
	citecolor=dark-red,
	urlcolor=dark-green,
	linktoc=all
}

\usepackage{geometry} 					
\numberwithin{equation}{section} 

\titleformat{\section}[block]{\Large\bfseries\centering}{\thesection}{1em}{} 
\titleformat{\subsection}[block]{\bfseries}{\thesubsection}{1em}{} 

\definecolor{dark-gray}{gray}{0.20}
\definecolor{gray}{gray}{0.30}
\definecolor{light-gray}{gray}{0.80}
\definecolor{dark-red}{rgb}{0.7,0,0}
\definecolor{dark-green}{rgb}{0.1,0.4,0}
\definecolor{dark-blue}{rgb}{0.3,0.3,0.7}
\definecolor{light-blue}{rgb}{0.8,0.8,1}
\definecolor{cardinal}{rgb}{0.6,0,0}
\definecolor{darkgreen}{rgb}{0,0.5,0}
\definecolor{golden}{rgb}{0.92, 0.7, 0}
\definecolor{midnight}{rgb}{0, 0, 0.5}
\definecolor{darkblue}{rgb}{0.2, 0, 0.8}
\definecolor{forestgreen}{rgb}{0.13, 0.55, 0.13}

\def\cH{{\cal H}}
\def\cI{{\cal I}}
\def\cJ{{\cal J}}
\def\cK{{\cal K}}
\def\cM{{\cal M}}
\def\cN{{\cal N}}

\def\cP{{\cal P}}
\def\cQ{{\cal Q}}
\def\cR{{\cal R}}
\def\cS{{\cal S}}

\def\Tr{{\rm Tr}\,}

\def\al{\alpha}
\def\ald{{\dot{\alpha}}}
\def\be{\beta}
\def\bed{{\dot{\beta}}}

\def\cW{{\cal W}}
\def\cL{{\cal L}}

\newcommand{\dd}{\mathrm{d}}
\newcommand{\e}{\mathrm{e}}
\newcommand{\f}[2]{\frac{#1}{#2}}

\newcommand\Tstrut{\rule{0pt}{2.6ex}}         

\def\SL{{\rm SL}}

\def\diag{{\rm diag}}

\def\SO{{\rm SO}}
\def\U{{\rm U}}
\def\SU{{\rm SU}}

\def\USp{{\rm USp}}
\def\Spin{{\rm Spin}}

\def\sl{\mathfrak{sl}}

\def\so{\mathfrak{so}}
\def\su{\mathfrak{su}}
\def\gl{\mathfrak{gl}}
\def\osp{\mathfrak{osp}}
\def\sp{\mathfrak{sp}}
\def\uu{\mathfrak{u}}

\title{\fontsize{20pt}{24pt}\selectfont\textbf{Comments on chiral algebras and $\Omega$-deformations}\vspace{3mm}}

\author{Nikolay Bobev$^{\Omega}$, Pieter Bomans$^{\Omega,\epsilon_1,\epsilon_2}$, and Fri\dh rik Freyr Gautason$^{\Omega,\epsilon_3}$\\[5mm]
	\normalsize $^\Omega$Instituut voor Theoretische Fysica, K.U. Leuven\\
	\normalsize Celestijnenlaan 200D, BE-3001 Leuven, Belgium\\[2mm]
	\normalsize $^{\epsilon_1}$Universit\`a di Padova, Dipartimento di Fisica e Astronomia\\
	\normalsize via Marzolo 8, 35131 Padova, Italy\\[2mm]
	\normalsize $^{\epsilon_2}$INFN, Sezione di Padova\\
	\normalsize via Marzolo 8, 35131 Padova, Italy\\[2mm]
	\normalsize $^{\epsilon_3}$University of Iceland, Science Institute\\
	\normalsize Dunhaga 3, 107 Reykjav\'ik, Iceland\\[3mm]
	\texttt{\small\href{mailto:nikolay.bobev@kuleuven.be}{nikolay.bobev@kuleuven.be}, \href{mailto:ffg@hi.is}{ffg@hi.is}, \href{mailto:pieter.bomans@unipd.it}{pieter.bomans@pd.infn.it}}\\
}

\date{}

\begin{document}  
	
\maketitle
	
\begin{abstract}
	\noindent Every six-dimensional $\cN=(2,0)$ SCFT on $\mathbf{R}^6$ contains a set of protected operators whose correlation functions are controlled by a two-dimensional chiral algebra. We provide an alternative construction of this chiral algebra by performing an~$\Omega$-deformation~of a topological-holomorphic twist of the $\cN=(2,0)$ theory on $\mathbf{R}^6$ and restricting to the cohomology of a specific supercharge. In addition, we show that the central charge of the chiral algebra can be obtained by performing equivariant integration of the anomaly polynomial of the six-dimensional theory. Furthermore, we generalize this construction to include orbifolds of the $\mathbf{R}^4$ transverse to the chiral algebra plane. 
\end{abstract}
	
\clearpage
	
\tableofcontents
	
\newpage	
	
\section{Introduction}

The operator product expansion (OPE) is a cornerstone of non-perturbative quantum field theory \cite{Wilson:1969zs,Polyakov:1974gs}. Unfortunately, it is prohibitively hard to delineate the detailed properties of the OPE for general QFTs. To make progress, it is fruitful to focus on QFTs with additional symmetries. Indeed, the OPE is best understood and most studied in the context of conformal field theory due to its better convergence properties and the tighter constraints imposed by conformal symmetry. For some two-dimensional CFTs, the associativity of the OPE combined with the powerful constraints from the Virasoro algebra can even lead to the exact solution for the correlation functions of local operators \cite{Belavin:1984vu}. This in turn led to a fruitful interplay between the mathematics of infinite-dimensional algebras and the physics of two-dimensional CFTs. However, it is not immediately clear whether these vertex operator algebras, or chiral algebras as we refer to them here, have any applications for QFTs in more than two dimensions. Supersymmetric QFTs present another class of examples where the properties of the OPE can be studied in greater detail. Indeed, employing ``the power of holomorphy'' \cite{Seiberg:1994bp} it was understood that the OPE of some protected operators in QFTs with at least four supercharges exhibits a special algebraic structure known as the chiral ring \cite{Lerche:1989uy}. 

These two techniques for gaining calculational insight into the properties of the OPE were nicely combined in \cite{Beem:2013sza,Beem:2014kka} where it was shown how non-trivial chiral algebras arise from SCFTs in 4d and 6d by restricting to the cohomology of a certain combination of Poincar\'e and conformal supercharges. The results of \cite{Beem:2013sza,Beem:2014kka} imply that every superconformal field theory containing $\mathfrak{psl}(2|2)$ as a subalgebra of its symmetry algebra contains a protected subset of operators with an interesting OPE structure. To determine this set of operators one has to study the cohomology of a specific nilpotent supercharge of the schematic form $\mathbbmtt{Q}=\mathcal{Q}+\mathcal{S}$ which is a linear combination of Poincar\'e and conformal supercharges. The non-trivial OPE of interest is then formed by local operators in the cohomology of this supercharge. For even-dimensional SCFTs on $\mathbf{R}^d$ with (at least) 6d $\mathcal{N}=(2,0)$, 4d $\cN=2$, and 2d $\mathcal{N}=(0,4)$ symmetry this cohomology results in a chiral algebra on a two-dimensional plane in $\mathbf{R}^d$.\footnote{For 3d $\cN=4$ SCFTs this cohomological construction leads to the OPE of operators on a real line \cite{Chester:2014mea}.} The correlation functions of these protected operators have a specific dependence on the holomorphic coordinate on the chiral algebra plane which is determined by their $\su(2)$ R-symmetry quantum numbers. These chiral algebras have been the subject of intense recent interest and have led to a plethora of new insights into the physics of 4d $\cN=2$ and 6d $\cN=(2,0)$ SCFTs, see \cite{Lemos:2020pqv} for a recent review and a comprehensive list of references.

Yet another way to tame the OPE in the context of supersymmetric QFTs is to employ the topological twist of Witten \cite{Witten:1988ze}. The basic idea behind this technique is to use the R-symmetry of a given supersymmetric QFT in flat space in order to modify the Lorentz transformations of operators in the theory in such a way that the resulting QFT is topological in nature and can be put on general curved manifolds. More generally, this method and its generalizations can be used in conjunction with supersymmetric localization to obtain non-trivial results for the path integral of the given supersymmetric QFT, see \cite{Pestun:2016zxk} for a recent comprehensive review. An important application of these ideas in the context of 4d $\cN=2$ QFTs is the $\Omega$-deformation \cite{Nekrasov:2002qd,Nekrasov:2003rj} which equips the QFT on $\mathbf{R}^4$ with two parameters, usually called $\epsilon_1$ and $\epsilon_2$, associated with rotations in the two orthogonal planes. This construction generalizes the Donaldson-Witten topological twist and by using supersymmetric localization it was shown that the path integral of the $\Omega$-deformed SYM theory on $\mathbf{R}^4$ reproduces the Seiberg-Witten prepotential in the $\epsilon_{1,2} \to 0 $ limit. Another important application of the $\Omega$-deformation arises in the AGT correspondence which relates observables in 4d $\cN=2$ QFTs of class $\mathcal{S}$, arising from M5-branes wrapped on a punctured Riemann surface \cite{Gaiotto:2009we}, defined on $\mathbf{R}^{4}_{\epsilon_{1},\epsilon_2}$ with correlation functions of the Liouville \cite{Alday:2009aq} (or more general Toda \cite{Wyllard:2009hg}) CFT defined on the Riemann surface, see \cite{LeFloch:2020uop}. Given these developments it is natural to ask whether the chiral algebras associated with 4d $\cN=2$ \cite{Beem:2013sza} and 6d $\cN=(2,0)$ SCFTs \cite{Beem:2014kka} can be obtained through a topological twist accompanied by an $\Omega$-deformation. In this paper we address this question.
		
Indeed, in \cite{Oh:2019bgz,Jeong:2019pzg} (see also Section 6 of \cite{Dedushenko:2019yiw}) it was shown that the chiral algebra of \cite{Beem:2013sza} associated to a 4d $\cN=2$ SCFT can be obtained in an alternative way. The starting point is the topological-holomorphic twist of Kapustin \cite{Kapustin:2006hi} defined for any 4d $\cN=2$ SCFT on the product manifold $\Sigma \times C$. The twist is implemented by identifying the structure group of $C$ with the Cartan subalgebra of the $\su(2)$ R-symmetry and that of $\Sigma$ with the $\mathfrak{u}(1)$ R-symmetry. Focusing on operators preserved by the supercharge invariant under this twist leads to a theory that is topological on $\Sigma$ and holomorphic on $C$. The resulting OPE on $C$ however is trivial since the operators are free to move on the surface $\Sigma$ and thus cannot have a singular OPE leading to an interesting chiral algebra on $C$. This problem can be remedied by supplementing the topological-holomorphic twist with an $\Omega$-deformation. It was shown in  \cite{Oh:2019bgz,Jeong:2019pzg} that by considering the two-dimensional $\Omega$-background $\Sigma = \mathbf{R}^2_{\epsilon}$ and taking $C=\mathbf{R}^2$ one finds an invariant supercharge $Q_{\Omega}$ which has the same cohomology as the supercharge used in the chiral algebra construction of~\cite{Beem:2013sza}. Moreover, by considering explicit examples of Lagrangian 4d $\cN=2$ QFTs it was shown that supersymmetric localization with respect to $Q_{\Omega}$ leads to precisely the same chiral algebra as in \cite{Beem:2013sza}. An important ingredient in the analysis of \cite{Oh:2019bgz,Jeong:2019pzg} is to show that the 4d $\cN=2$ SCFT on $\mathbf{R}^2_{\epsilon} \times \mathbf{R}^2$ is not deformed, i.e. it has the same path integral as the theory on $\mathbf{R}^4$. This establishes a direct relation between the two constructions of the chiral algebra.

Encouraged by this result, it is natural to wonder whether one can obtain the 6d $\cN=(2,0)$ SCFT chiral algebra of \cite{Beem:2014kka} through a similar $\Omega$-deformed topological twist. The purpose of this paper is to show that this is indeed possible and to outline the details of the construction. As a starting point, we revisit the Donaldson-Witten twist of 4d $\mathcal{N}=2$ QFTs and apply $\Omega$-deformation as described by Nekrasov \cite{Nekrasov:2002qd}. Next, we explicitly identify the combination of supercharges, $Q_{\Omega}$, invariant under this deformation. We then show that when the 4d $\mathcal{N}=2$ theory is conformal and one takes $\epsilon_1=\epsilon_2$ the supercharge $Q_{\Omega}$ becomes a linear combination of Poincar\'e and conformal supercharges of the theory on $\mathbf{R}^4$. This in turn implies that for this special choice of $\Omega$-deformation, the SCFT on $\mathbf{R}^4$ is not deformed and the effect of the $\Omega$-deformation is to select an appropriate set of protected operators in the cohomology of $Q_{\Omega}$ that form an interesting cohomology. Equipped with this understanding we then study a topological-holomorphic twist of the 6d $\cN=(2,0)$ SCFT on a manifold of the form $X_4\times C$. The twist is defined by considering an $\su(2)\times \uu(1)$ subalgebra of the $\so(5)$ R-symmetry and then performing a twist with $\su(2)$ on $X_4$ and with $\uu(1)$ on $C$. Choosing $X_4=\mathbf{R}^{4}_{\epsilon_{1},\epsilon_2}$ and $C=\mathbf{R}^2$ and specializing to $\epsilon_1=\epsilon_2$ relates this to the twist and $\Omega$-deformation described above and selects a specific $Q_{\Omega}$ supercharge which we construct explicitly. We then show that this is precisely the same supercharge as the one used in the chiral algebra construction of \cite{Beem:2014kka} establishing an equivalence between the two chiral algebras. Ideally one would then like to use this result and employ supersymmetric localization of the 6d $\cN=(2,0)$ SCFT of type $\mathfrak{g}=\{A_N,D_N,E_{6,7,8}\}$ with respect to this $Q_{\Omega}$ supercharge to derive the well-supported conjecture of \cite{Beem:2014kka} that the resulting chiral algebra is the $\mathcal{W}_{\mathfrak{g}}$ algebra. Unfortunately, this is hard to do due to the lack of Lagrangian description of the 6d $\cN=(2,0)$ SCFT. In the absence of such an explicit derivation of the chiral algebra we resort to a more indirect method to check our results. Using the results of \cite{Alday:2009qq}, we show that an equivariant integration of the anomaly polynomial of the 6d $\cN=(2,0)$ SCFT on $\mathbf{R}^{4}_{\epsilon_{1},\epsilon_2} \times\mathbf{R}^2$ leads to the central charge of the $\mathcal{W}_{\mathfrak{g}}$ algebra in the limit $\epsilon_1=\epsilon_2$. Moreover, we lend further support for the validity of the construction in \cite{Oh:2019bgz,Jeong:2019pzg} by performing a similar equivariant integration of the anomaly polynomial of 4d $\cN=2$ SCFTs in order to recover the expected central charge and flavor symmetry current level of \cite{Beem:2013sza}. Our results also admit a natural generalization where one considers $\mathbf{Z}_k$ orbifolds of the $X_4=\mathbf{R}^{4}_{\epsilon_{1},\epsilon_2}$ space transverse to the chiral algebra plane $C$. Inspired by the results in \cite{Belavin:2011pp,Nishioka:2011jk,Bonelli:2011jx}  we conjecture that the resulting chiral algebra is related to the $k^{\rm th}$ para-$\cW_{\mathfrak{g}}$ algebra.

In the next section we begin our exploration by discussing the $\Omega$-deformation of topologically twisted 4d $\cN=2$ SCFTs. In Section~\ref{sec:6dchiral} we show how this can be used to formulate a topological-holomorphic twist accompanied by an $\Omega$-deformation for the 6d $\cN=(2,0)$ SCFT and show how the chiral algebra of \cite{Beem:2014kka} arises from this construction. In Section~\ref{sec:centralc} we use the equivariant integration of the anomaly polynomial to derive the central charges of the 4d $\cN=2$ and 6d $\cN=(2,0)$ chiral algebras. We conclude with a summary and a discussion of some open problems in Section~\ref{sec:discussion}. In Appendices ~\ref{App:bispin}, \ref{app:42}, and \ref{app:20} we collect our conventions and some details on the 4d and 6d superconformal algebras used in the main text. In Appendix~\ref{App:OhYagi} we clarify some aspects of the calculation of supercharges in the construction of \cite{Oh:2019bgz,Jeong:2019pzg}. Finally, Appendix~\ref{App:N4SCA} is devoted to a derivation of the chiral algebra associated to 2d $\cN=(0,4)$ SCFT.

\section{$\Omega$-deformation of topologically twisted 4d $\cN=2$ SCFTs}
\label{sec:4dN2}

Let us start by recalling some basic facts about the four-dimensional conformal algebra $\so(5,1)$.\footnote{We will be often cavalier about the distinction between Lorentzian and Euclidean conformal algebras. For our purposes, one can work with the complexified version of the (super)conformal algebras and restrict to a real form when needed. } It is generated by translations, special conformal transformations, rotations, and dilatations with respective generators
\begin{equation}
	\cP_{\al\ald}\,,\qquad \cK^{\ald\al}\,,\qquad {\cM_\al}^\be\,,\qquad {\cM^\ald}_\bed\,,\qquad \cH\,,
\end{equation}
where $\al,\be = \pm$ and $\ald,\bed = \dot{\pm}$ are spinorial indices for Weyl spinors and their complex conjugates. By adding eight Poincar\'e supercharges $\cQ_\al^\cI$, $\widetilde{\cQ}_{\cI\ald}$ and eight conformal supercharges $\cS_\cI^\al$, $\widetilde{\cS}^{\cI\ald}$ to this algebra we obtain the 4d $\cN=2$ superconformal algebra $\su(4^\star|2)$. These supercharges are rotated into each other by the $\U(2)_\cR \simeq \SU(2)_R \times \U(1)_r$ R-symmetry with generators ${\cR^\cI}_\cJ$ where $\cI,\cJ = 1,2$ are indices in the fundamental of $\U(2)_\cR$. The commutation relations of this algebra are summarized in Appendix \ref{app:42}. 

We are interested in studying $\cN=2$ SCFTs defined on an arbitrary spin manifold $X_4$ with Riemannian metric $g_{\mu\nu}$. In general, the holonomy group of the manifold is $\Spin(4)$ but we will also consider manifolds of the form $X_4 = \Sigma \times \Sigma_\perp$, where $\Sigma$ and $\Sigma_\perp$ are two surfaces. In this case, the holonomy reduces to $\SO(2)_\Sigma\times \SO(2)_{\Sigma_\perp}$ and we will always choose a local frame such that the generators of the holonomy group are
\begin{equation}
	\begin{aligned}
		\cL_\Sigma &= \f12\left({{\cM}_+}^+-{{\cM}_-}^- + {{\cM}^{\dot{+}}}_{\dot{+}}-{{\cM}^{\dot{-}}}_{\dot{-}}\right)\,,\\
		\cL_{\Sigma_\perp} &= \f12 \left({{\cM}_+}^+-{{\cM}_-}^- - {{\cM}^{\dot{+}}}_{\dot{+}}+{{\cM}^{\dot{-}}}_{\dot{-}}\right)\,.
	\end{aligned}
\end{equation}
We denote the eigenvalues of these generators as $h_\Sigma$ and $h_{\Sigma_\perp}$. The generators of the Cartan of the two $\SU(2)$'s in $\Spin(4)\simeq \SU(2)_+ \times \SU(2)_-$ on the other hand are given by $\cL_\Sigma+\cL_{\Sigma_\perp}$ and $\cL_\Sigma-\cL_{\Sigma_\perp}$ respectively. In Table~\ref{tab:charges42} we present the Poincar\'e supercharges together with their quantum numbers.
\begin{table}[!htb]
	\centering
	\begin{tabular}{l|cccccccc}
		& $\cQ_+^1$ & $\cQ_-^1$ & $\cQ_+^2$ & $\cQ_-^2$& $\widetilde{\cQ}_{1\dot{+}}$& $\widetilde{\cQ}_{1\dot{-}}$ & $\widetilde{\cQ}_{2\dot{+}}$ & $\widetilde{\cQ}_{2\dot{-}}$\\[1mm]
		\hline
		$\rule[-1pt]{0pt}{0.6cm} h_\Sigma$ & $\f12$ & $-\f12$& $\f12$& $-\f12$& $-\f12$& $\f12$& $-\f12$& $\f12$\\[1mm]
		$h_{\Sigma_\perp}$ & $\f12$ & $-\f12$& $\f12$& $-\f12$& $\f12$& $-\f12$& $\f12$& $-\f12$\\[1mm]
		\hline
		$\rule[-1pt]{0pt}{0.6cm}R$ & $\f12$ & $\f12$& $-\f12$& $-\f12$& $-\f12$& $-\f12$& $\f12$& $\f12$\\[1mm]
		$r$ & $\f12$ & $\f12$& $\f12$& $\f12$& $-\f12$& $-\f12$& $-\f12$& $-\f12$\\[1mm]
	\end{tabular}
	\caption{$\cN=2$ supercharges and their quantum numbers. $R$ and $r$ denote the charges under $\U(1)_R \subset \SU(2)_R$ and $\U(1)_r$ respectively. See Equation~\eqref{eq:rR4ddef} for our conventions.}
	\label {tab:charges42}
\end{table}

In general when one places an $\cN=2$ theory on a curved space $X_4$ all supersymmetry is broken. However, there are various ways to preserve a fraction of the original supersymmetries; one of them is by performing a (partial) topological twist. The twisting procedure we are interested in, originally introduced by Witten in \cite{Witten:1988ze}, consists of redefining the Lorentz representations of the fields in order to construct a conserved topological supercharge. This is achieved by turning on a background vector field that couples to the $\SU(2)_R$ current and identifying it with the $\SU(2)_-$ spin connection. More concretely, this means that we identify the $\SU(2)_R$ indices $\cI$ and $\cJ$ with the indices $\ald$ and $\bed$. We define the twisted rotation generators ${\cM^{\prime\ald}}_{\bed}$ as follows:
\begin{equation}\label{Wittentwist}
	{\cM^{\prime\ald}}_{\bed} = {\cM^{\ald}}_{\bed} - \left({\cR^{\ald}}_\bed-\f12{\delta^{\ald}}_\bed \cR^{\dot{\eta}}\,_{\dot{\eta}}\right)\,.
\end{equation}
In the twisted theory, the group $\cK^\prime = \SU(2)_+\times \SU(2)_-^\prime$ acts as the new rotation group on the tangent space of the four-manifold, where $\SU(2)_-^\prime$ is generated by $\cM^\prime$. The transformation properties of the eight Poincar\'e supercharges under the new rotation group are given by 
\begin{equation}\label{eq:4dQdecomp}
	(\mathbf{1},\mathbf{2},\mathbf{2}_R)_{\frac{1}{2}} \to (\mathbf{1},\mathbf{1}^\prime)_{\frac{1}{2}} \oplus (\mathbf{1},\mathbf{3}^\prime)_{\frac{1}{2}}\,,\qquad (\mathbf{2},\mathbf{1},\mathbf{2}_R)_{-\frac{1}{2}} \to (\mathbf{2},\mathbf{2}^\prime)_{-\frac{1}{2}}\,.
\end{equation}
Importantly, under the new rotation group the supercharge
\begin{equation}\label{Qtop}
	Q \equiv \epsilon^{\ald\bed}{\widetilde{\cQ}}_{\ald\bed} = \widetilde{\cQ}_{\dot{-}\dot{+}}-\widetilde{\cQ}_{\dot{+}\dot{-}}\,,
\end{equation}
corresponding to $(\mathbf{1},\mathbf{1}^\prime)_{\frac{1}{2}}$ in \eqref{eq:4dQdecomp}, is a nilpotent scalar, i.e. it obeys
\begin{equation}\label{eq:Wtwscalar}
	[ {\cM_{\al}}^\be , Q ] = 0\,, \qquad 	[ {\cM^{\prime\ald}}_\bed , Q ] = 0\,, \qquad Q^2 = 0\,. 
\end{equation}
In addition to this scalar supercharge we can define the vector supercharge $G_{\al{\ald}} = i{\cQ}_{\ald\al}$, or equivalently, $G_\mu = i(\bar{\sigma}_\mu)^{\ald\al} {\cQ}_{\ald\al}$, which corresponds to $(\mathbf{2},\mathbf{2}^\prime)_{-\frac{1}{2}}$ in  \eqref{eq:4dQdecomp}. Using the commutation relations of the supersymmetry algebra it is easy to show that 
\begin{equation}\label{Ptop}
	\{Q , G_{\al\ald}\} = i\cP_{\al\ald}\,.
\end{equation}
By considering the cohomology with respect to the scalar supercharge $Q$ one obtains a topological field theory. Indeed, from \eqref{Ptop} it is clear that all translations are $Q$-exact. The $Q$-cohomology is graded by $r$. Note that for $X_4=\mathbf{R}^4$ the topological twist does not change or deform the  QFT. It is merely a relabeling of the fields in the 4d $\cN=2$ theory and can be used to select a subset of operators which forms a topological field theory when passing to the cohomology of $Q$.

We are not interested in this Donaldson-Witten topological twist itself but rather in a deformation of it. To formulate this deformation let us consider the topological twist on a four-manifold $X_4$ that has an isometry generated by a Killing vector field $V$. In \cite{Nekrasov:2002qd,Nekrasov:2003rj} a deformation of the twisted theory with respect to $V$ was introduced where the topological supercharge $Q$ is replaced with a new supercharge $Q_\Omega$ such that 
\begin{equation}
	Q_{\Omega}^2 = \cL_V\,,
\end{equation}
where $\cL_V$ is the conserved charge corresponding to $V$ which acts on fields as the Lie derivative with respect to $V$.\footnote{More generally, we allow $V$ to be a complex linear combination of (commuting) Killing vector fields.} Since $Q_{\Omega}$ is no longer nilpotent, the appropriate framework to study the cohomological theory is equivariant cohomology and for consistency, we have to restrict ourselves to $V$-invariant operators and states.

For example, if the 4d $\mathcal{N}=2$ theory is formulated on flat space, the topological twist does not deform the original QFT and thus the vector supercharge discussed below \eqref{eq:Wtwscalar} is still a symmetry generator. If one then defines the $\Omega$-deformation of $\mathbf{R}^4$ with respect to the covariantly constant Killing vector  $V = V^\mu\partial_\mu$ then $\iota_V G = V^\mu G_\mu$ is a conserved charge and we can write the $\Omega$-deformed supercharge as
\begin{equation}\label{Qdef}
	Q_\Omega = Q + \iota_V G\,.
\end{equation} 
It can be checked explicitly that this indeed satisfies $Q_\Omega^2 = \cL_V$. 

As emphasized in \cite{Nekrasov:2002qd,Nekrasov:2003rj} the $\Omega$-deformation can be defined in the more general situation when $X_4$ is not flat and $V$ is not covariantly constant. In this case, the supersymmetry transformations of the various fields in the original 4d $\mathcal{N}=2$ QFT have to be modified and the action of the theory has to be supplemented by extra terms such that the deformed theory is invariant under $Q_\Omega$. We therefore conclude that generically the $\Omega$-deformation results in a genuine deformation of the 4d $\mathcal{N}=2$ QFT. In special cases, however, the 4d $\mathcal{N}=2$ QFT might already be invariant under the $\Omega$-deformed supercharge. Specifically, this happens when the deformed supercharge $Q_\Omega$ is a symmetry the undeformed twisted theory. In such a situation the $\Omega$-deformation does not truly deform the 4d $\mathcal{N}=2$ QFT but only changes the cohomology we consider. We now discuss precisely such a scenario which is realized when the 4d $\mathcal{N}=2$ theory is conformal.

\subsection{$\cN=2$ SCFTs on $\mathbf{R}^4_{\epsilon_{1},\epsilon_{2}}$}
\label{subsec:4dN2Omega}

Now suppose that the original theory we consider is an $\cN=2$ SCFT defined on $\mathbf{R}^4=\mathbf{R}^2\times \mathbf{R}^2$. As discussed above for some choices of the vector $V$, the quantity $\iota_V G$ is a linear combination of supercharges of the original SCFT hence the $\Omega$-deformation does not truly deform the twisted theory. For example, this is clearly true for $V$ being one of the generators of translations. We will consider the more interesting case when $V$ is a linear combination of rotations in the two orthogonal planes of $\mathbf{R}^4$, generated by $\cL_\Sigma$ and $\cL_{\Sigma_\perp}$. The vector $V$ corresponding to this rotation can be written as
\begin{equation}
	V = 2\Omega^{\mu\nu}x_\nu\partial_\mu = 2\epsilon_{1} x_{[2}\f{\partial}{\partial x^{1]}} + 2\epsilon_{2} x_{[4}\f{\partial}{\partial x^{3]}}\,,
\end{equation}
where the matrix $\Omega^{\mu\nu}$ is defined as
\begin{equation}
	\Omega^{\mu\nu} = \begin{pmatrix}
		0 & \epsilon_1 & 0 & 0 \\
		-\epsilon_1 & 0 & 0 & 0 \\
		0 & 0 & 0 & \epsilon_2 \\
		0 & 0 & -\epsilon_2 & 0
	\end{pmatrix}\,.
\end{equation}
To describe the contraction of $V$ with the one-form supercharge $G$ it will prove useful to write the supersymmetry and superconformal generators as three-dimensional integrals of local supercurrents:
\begin{align}
	{\cQ^{\dot{\alpha}}}_{\beta} &= \int \dd^3 x \, {\mathcal{G}^{\dot{\alpha}}}_{\beta} \,,&
	{\cS_{\dot{\alpha}}}^{\beta} &= \int \dd^3 x \,  \left(\sigma^\mu\right)^{\beta\dot{\beta}}x_\mu \bar{\mathcal{G}}_{\dot{\alpha}\dot{\beta}}\,,\\
	\widetilde{\cQ}_{\dot{\alpha}\dot{\beta}} &= \int \dd^3 x \, \bar{\mathcal{G}}_{\dot{\alpha}\dot{\beta}} \,,&
	\widetilde{\cS}^{\dot{\alpha}\dot{\beta}} &= \int \dd^3 x \,  \left(\bar{\sigma}^\mu\right)^{\dot{\beta}\be} x_\mu {\mathcal{G}^{\dot{\alpha}}}_{\beta} \,.
\end{align}
In terms of these supercurrents, the one-form supercharge takes the form
\begin{equation}\label{eq:GDWdef}
	G = i(\bar{\sigma}_\mu)^{\dot{\alpha}\alpha} \cQ_{\dot{\alpha}\alpha} \dd x^\mu = i\left(\int \dd^3 x \, (\bar{\sigma}_\mu)^{\dot{\alpha}\alpha} \mathcal{G}_{\dot{\alpha}\alpha} \right)\dd x^\mu\,.
\end{equation}
A natural definition for the contraction of this one-form supercharge with the vector $V$ is then given by
\begin{equation}\label{eq:VGcontract4dN2}
	\iota_V G = 2i\int \dd^3 x \, \Omega^{\mu\nu}x_\nu (\bar{\sigma}_\mu)^{\dot{\alpha}\alpha} \mathcal{G}_{\dot{\alpha}\alpha} \,.
\end{equation}
For generic values of the deformation parameters, $\epsilon_1$ and $\epsilon_2$, the integral in \eqref{eq:VGcontract4dN2} is not a conserved charge of the original 4d $\mathcal{N}=2$ SCFT. Thus for generic $\epsilon_1$ and $\epsilon_2$ the $\Omega$-deformation genuinely deforms the twisted 4d $\mathcal{N}=2$ SCFT. However, in the special case when $\epsilon_{1} = \epsilon_2=\f{\epsilon}{2}$ one finds that 
\begin{equation}
	\iota_V G\Big|_{\epsilon_{2}=\epsilon_{1}} = \, \epsilon \left(\tilde{\cS}^{\dot{+}\dot{-}}+\tilde{\cS}^{\dot{-}\dot{+}}\right)\,,
\end{equation}
such that, using \eqref{Qtop} and \eqref{Qdef}, the $\Omega$-deformed supercharge can be written as 
\begin{equation}\label{Qom}
	Q_\Omega = \tilde{\cQ}_{\dot{-}\dot{+}}-\tilde{\cQ}_{\dot{+}\dot{-}} + \epsilon \left( \tilde{\cS}^{\dot{+}\dot{-}}+\tilde{\cS}^{\dot{-}\dot{+}} \right)\,,
\end{equation}
which satisfies
\begin{equation}\label{eq:Qomega4dN2}
	Q_\Omega^2 = 2\,\epsilon \left({\cM^{\prime\dot{+}}}_{\dot{+}}-{\cM^{\prime\dot{-}}}_{\dot{-}}\right) = \cL_V\,.
\end{equation}
From now on we specify to $\epsilon_{1}=\epsilon_2=\f{\epsilon}{2}$ and denote by $Q_\Omega$ the corresponding $\Omega$-deformed supercharge. It is clear that $Q_\Omega$ is written entirely in terms of the generators of the 4d $\mathcal{N}=2$ superconformal algebra and therefore we do not have to modify the 4d $\mathcal{N}=2$ SCFT to ensure invariance under $Q_\Omega$.\footnote{Similarly, we can choose to perform the Donaldson-Witten twist using the $\SU(2)_+$ instead of $\SU(2)_-$ in \eqref{Wittentwist}. We can then repeat all the steps above and find that the $\Omega$-deformation with $\epsilon_{1}=-\epsilon_2$ does not deform the 4d $\mathcal{N}=2$ SCFT.}

Next, we want to characterize the cohomology of local operators with respect to $Q_\Omega$. Due to \eqref{eq:Qomega4dN2} these operators must be invariant under the twisted rotation $\cL_V$. Indeed, this is expected based on the framework of equivariant cohomology with respect to the vector $V$. Notice that the equivariant $Q_\Omega$-cohomology is isomorphic for all values of $\epsilon\neq 0$. We also note that the Poincar\'e and conformal supercharges on the right-hand side of \eqref{Qom} have different charges under the $\U(1)_r$ symmetry. To remedy this we can assign degree $2$ to $\epsilon$ so that the cohomology of interest is still graded by the $\U(1)_r$ charge.

Notice that we can write $Q_\Omega$ as $Q_\Omega = Q_\Omega^{(1)} + Q_\Omega^{(2)}$ where
\begin{equation}\label{Q12}
	Q_\Omega^{(1)} = - \widetilde{\cQ}_{\dot{+}\dot{-}}+\epsilon\widetilde{\cS}^{\dot{-}\dot{+}}\,,\qquad \qquad 	Q_\Omega^{(2)} = \widetilde{\cQ}_{\dot{-}\dot{+}}+\epsilon\widetilde{\cS}^{\dot{+}\dot{-}}\,.
\end{equation}
Importantly, we find that $\left(Q_\Omega^{(1)}\right)^2=\left(Q_\Omega^{(2)}\right)^2=0$. This implies that both $Q_\Omega^{(1)}$ and $Q_\Omega^{(2)}$ define their own cohomology. Since twisted rotations are $Q_\Omega$-exact as well as $Q_\Omega^{(1)}$- and $Q_\Omega^{(2)}$-exact, to study these cohomologies it suffices to consider operators inserted at the origin invariant under $\cL_V$. A local operator inserted at the origin is a harmonic representative of a $Q_\Omega$-cohomology class if it is annihilated by both $Q_\Omega$ and $Q_\Omega^\dagger$. This happens if and only if it is annihilated by $\left\{Q_\Omega,Q_\Omega^\dagger\right\}$. Given that the cohomology of $Q_\Omega$ is independent of $\epsilon$ we are free to fix $|\epsilon|=1$. One then finds that the three cohomologies are related due to the following identity\footnote{A related identity can be shown to hold for any $\epsilon$ when acting on operators annihilated by \eqref{Q12commutator}.}
\begin{equation}
	\begin{aligned}
			\left\{Q_\Omega,Q_\Omega^\dagger\right\} &= 2 \left\{Q_\Omega^{(1)},\left(Q^{(1)}_\Omega\right)^\dagger\right\} = 2 \left\{Q^{(2)}_\Omega,\left(Q^{(2)}_\Omega\right)^\dagger\right\}\\&= 2\left(\cH +2r\right)\,,
	\end{aligned}
\end{equation}
which in terms of the four-dimensional quantum numbers reduces to the condition $\Delta~+~2r~=~0$, where $\Delta$ is the eigenvalue of $\cH$. We thus find that the equivariant cohomology of $Q_\Omega$ is the same as that of $Q_\Omega^{(1)}$ and $Q_\Omega^{(2)}$ supplemented with the condition that the operators in cohomology should be annihilated by the generator
\begin{equation}\label{Q12commutator}
	\left\{Q_\Omega^{(1)},Q_\Omega^{(2)}\right\} = \epsilon \left({\cM^{\prime\dot{+}}}_{\dot{+}}-{\cM^{\prime\dot{-}}}_{\dot{-}}\right) = \epsilon\left({\cM^{\dot{+}}}_{\dot{+}}-{\cM^{\dot{-}}}_{\dot{-}}-2R\right)\,.
\end{equation}
These operators have quantum numbers that obey $h_{\Sigma}-h_{\Sigma_\perp} = 2R$. It can be checked that operators obeying such constraints on their quantum numbers belong to short and semi-short multiplets of the 4d $\cN=2$ superconformal algebra, see for example \cite{Cordova:2016emh}.\footnote{Note that our normalizations for the superconformal algebra quantum numbers differ by various factors of 2 from the conventions in \cite{Cordova:2016emh}.}

We stress again that for $\epsilon_1=\epsilon_2$ the topologically twisted and $\Omega$-deformed 4d $\cN=2$ SCFT is in fact not deformed and has the same Lagrangian and symmetries as the theory on $\mathbf{R}^4$. This bodes well with previous results in the literature. The $\Omega$-deformation was studied from the perspective of 4d $\cN=2$ rigid conformal supergravity in \cite{Klare:2013dka} where it was shown that only for $\epsilon_1\neq\epsilon_2$  the supergravity background is non-trivial and the Lagrangian of the SCFT is deformed. This was also confirmed in \cite{Bobev:2019ylk} where the holographic dual of 4d $\cN=2$ SCFTs on $\Omega$-deformed $\mathbf{R}^4$ was studied.

We end our discussion of the $\Omega$-deformation of 4d $\cN=2$ QFTs on $\mathbf{R}^4$ by noting that this setup has been studied extensively in the context of supersymmetric localization. The path integral of the QFT on this background can be reduced to an infinite sum of ordinary integrals over instanton moduli spaces
\begin{equation}
	\mathcal{Z}(\vec{a},\tau;\epsilon_{1,2}) = \mathcal{Z}^{\rm pert}(\vec{a},\tau;\epsilon_{1,2})\left(1+\sum_{k=1}^\infty q^k \mathcal{Z}^{(k)}(\vec{a},\tau;\epsilon_{1,2})\right)\,,
\end{equation}
where $q=\exp(2\pi i\tau)$, $\tau$ is the complexified gauge coupling, $\vec{a}$ are the Coulomb branch moduli, and $\mathcal{Z}^{(k)}(\vec{a},\tau;\epsilon_{1,2})$ denotes the partition function in the $k$ instanton sector. The prefactor $\mathcal{Z}^{\rm pert}(\vec{a},\tau;\epsilon_{1,2})$ is explicitly known and given by the product of a tree-level contribution and a one-loop determinant factor. The instanton contributions were understood comprehensively in the seminal work \cite{Nekrasov:2002qd,Nekrasov:2003rj} where it was shown how to perform the integrals over the instanton moduli space of the gauge theory $\cM_G$ explicitly for general $k$. This approach allows to explicitly recover the Seiberg-Witten prepotential \cite{Seiberg:1994rs,Seiberg:1994aj} in the limit $\epsilon_{1,2} \rightarrow 0$. A prominent characteristic of the Nekrasov partition function is its relation to two-dimensional chiral algebras of the Liouville/Toda type. This is the basis of the AGT correspondence \cite{Alday:2009aq} and its generalizations, see for example \cite{Wyllard:2009hg}. In the context of these results, it is known that the Nekrasov partition function simplifies in various limits of the parameters $\epsilon_{1,2}$. The observation we made above that for 4d $\cN=2$ SCFTs the $\Omega$-deformation with $\epsilon_1=\epsilon_2$ does not deform the theory and thus preserves the full superconformal invariance may offer an insight into some of these simplifications.

\section{Chiral algebras from 6d $\cN=(2,0)$ SCFTs}
\label{sec:6dchiral}

We are now ready to apply the above results for topologically twisted 4d $\cN=2$ SCFTs in the context of the 6d $\cN=(2,0)$ theory and show how one can obtain the chiral algebra of \cite{Beem:2014kka}.
	
Before we start let us recall some basic facts about the 6d $\cN=(2,0)$ superconformal algebra. The 6d conformal algebra, $\so(7,1)$, is generated by translations, special conformal transformations, rotations, and dilatations, with respective generators
\begin{equation}
\cP_{ab}\,,\qquad \cK^{ab}\,,\qquad {\cM_a}^b\,,\qquad \cH\,.
\end{equation}
In addition to this the superconformal algebra contains sixteen Poincar\'e and conformal supercharges $\cQ_{Aa}$ and $\cS_A^a$ as well as the generators ${\cR}_{AB}$ of the R-symmetry group $\USp(4) \simeq \Spin(5)$. The $a,b,\dots=1,\dots,4$ indices transform in the fundamental representation of $\SU(4)\simeq \Spin(6)$ while the $A,B,\dots=1,\dots,4$ indices transform in the fundamental representation of $\USp(4)$. 
More details as well as the commutation relations of this superconformal algebra are summarized in Appendix \ref{app:20}.
	
In the following we choose a local frame such that an orthogonal basis for the Cartan of $\so(6)$ is given by the following generators of rotations in the three orthogonal planes in $\mathbf{R}^6$,
	\begin{equation}\label{rot}
	\begin{aligned}
	\cL_1 &\equiv \f12(\cM_1^1+\cM_2^2-\cM_3^3-\cM_4^4)\,,\\
	\cL_2 &\equiv \f12(\cM_1^1-\cM_2^2+\cM_3^3-\cM_4^4)\,,\\
	\cL_3 &\equiv \f12(\cM_1^1-\cM_2^2-\cM_3^3+\cM_4^4)\,.\\
	\end{aligned}
	\end{equation}
We denote the eigenvalues of these generators with $h_i$. In the following, we are interested in the 6d $\cN=(2,0)$ theory on a manifold of the form $X_4 \times C$. On this type of product manifold, the holonomy is reduced to
\begin{equation}
{\rm Spin}(4)\times{\rm Spin}(2) \simeq \SU(2)_+ \times \SU(2)_-\times \U(1)_C\,,
\end{equation}
The rotations on $C$ can be identified with $\cL_1$, while the Cartan of the holonomy group of $X_4$ is generated by $\cL_2$ and $\cL_3$. The eigenvalues of the Cartan of $\SU(2)_+\times\SU(2)_-$ are given by $h_2+h_3$ and $h_2-h_3$, respectively. An important role in the construction is played by the maximal subgroup of the R-symmetry $\SU(2)_R \times \U(1)_r \subset \USp(4)$ with generators $R_\pm$, $R$ and $r$ given by\footnote{Note that there is a choice when we specify this subgroup. Namely in \eqref{Rid} we are free to simultaneously change the definition of $R_+ \leftrightarrow -R_-$ and $R \rightarrow -R$. This will play an important role below. \label{ftn:SU2}}
\begin{equation}\label{Rid}
R_+ = \cR_{13}\,,\quad R_- = \cR_{24}\quad\,, R = -\f12\left(\cR_{14}-\cR_{23}\right)\,,\quad r = \f12\left(\cR_{14}+\cR_{23}\right)\,.
\end{equation}
In a slight abuse of notation the eigenvalues under the Cartan of this subalgebra, generated by $R$ and $r$, are denoted by $(R,r)$. In Table \ref{tab:charges20} we summarize the quantum numbers of the supercharges under the global symmetries.
	
The supercharges transform in the $(\mathbf{4}_+, \mathbf{4})$ of ${\rm Spin}(6)\times {\rm Spin}(5)$, where $\mathbf{4}_+$ is a positive chirality spinor of ${\rm Spin}(6)$. Under $\SU(2)_+ \times \SU(2)_- \times \SU(2)_R \times \U(1)_C \times \U(1)_r$ the supercharges decompose as
\begin{equation}\label{eq:6d20supercharges}
({\bf 4}_+,{\bf 4}) \to ({\bf 2},{\bf 1},{\bf 2})_{\f12,\f12} \oplus ({\bf 1},{\bf 2},{\bf 2})_{-\f12,\f12} \oplus ({\bf 2},{\bf 1},{\bf 2})_{\f12,-\f12} \oplus  ({\bf 1},{\bf 2},{\bf 2})_{-\f12,-\f12}\,.
\end{equation}

		\begin{table}[!htb]
		\centering
		\begin{tabular}{c|c|c|c}
			$\cQ_{Aa}$ & $h_1,h_2,h_3$ & $(R,r)$ & $h_1^\prime$ \\
			\hline
			$\cQ_{11}$ & $+,+,+$ & $(+,+)$ & $1$ \\
			$\cQ_{21}$ & $+,+,+$ & $(-,+)$ & $1$ \\
			$\cQ_{31}$ & $+,+,+$ & $(+,-)$ & $0$ \\
			$\cQ_{41}$ & $+,+,+$ & $(-,-)$ & $0$ \\
			\hline
			$\cQ_{12}$ & $+,-,-$ & $(+,+)$ & $1$ \\
			$\cQ_{22}$ & $+,-,-$ & $(-,+)$ & $1$ \\
			$\cQ_{32}$ & $+,-,-$ & $(+,-)$ & $0$ \\
			$\cQ_{42}$ & $+,-,-$ & $(-,-)$ & $0$ \\
		\end{tabular}
		\hspace{20mm}
		\begin{tabular}{c|c|c|c}
			$\cQ_{Aa}$ & $h_1,h_2,h_3$ & $(R,r)$ & $h_1^\prime$ \\
			\hline
			$\cQ_{13}$ & $-,+,-$ & $(+,+)$ & $0$ \\
			$\cQ_{23}$ & $-,+,-$ & $(-,+)$ & $0$ \\
			$\cQ_{33}$ & $-,+,-$ & $(+,-)$ & $-1$ \\
			$\cQ_{43}$ & $-,+,-$ & $(-,-)$ & $-1$ \\
			\hline
			$\cQ_{14}$ & $-,-,+$ & $(+,+)$ & $0$\\
			$\cQ_{24}$ & $-,-,+$ & $(-,+)$ & $0$ \\
			$\cQ_{34}$ & $-,-,+$ & $(+,-)$ & $-1$ \\
			$\cQ_{44}$ & $-,-,+$ & $(-,-)$ & $-1$ \\
		\end{tabular}
		\caption{The spins, R-charges and twisted spins of the Poincar\'e supercharges. All untwisted quantum numbers have magnitude $1/2$.}
		\label{tab:charges20}
	\end{table}
	
The first step to obtain a chiral algebra from a 6d $\cN=(2,0)$ theory is to twist the theory on $X_4\times C$. We choose a twist that results in a theory that is topological along $X_4$ and holomorphic along $C$. This twist is very similar to the twist studied by Kapustin in \cite{Kapustin:2006hi}, and used in \cite{Oh:2019bgz,Jeong:2019pzg} to obtain the chiral algebra associated to 4d $\cN=2$ SCFTs. We implement the twist by taking the following twisted rotation group
\begin{equation}
\cK^\prime = \SU(2)_+ \times \SU(2)^\prime_- \times \U(1)_C^\prime\,,
\end{equation}
where $\SU(2)^\prime_- = \diag(\SU(2)_-\times \SU(2)_R)$ and $\U(1)_C^\prime = \diag(\U(1)_C \times \U(1)_{r})$. Using \eqref{eq:6d20supercharges} we find that under $\cK^\prime$ the Poincar\'e supercharges transform as
\begin{equation}\label{twistrep}
(\mathbf{2},\mathbf{2})_{1}\oplus(\mathbf{2},\mathbf{2})_{0}\oplus(\mathbf{1},\mathbf{1})_{0}\oplus(\mathbf{1},\mathbf{3})_{0}\oplus(\mathbf{1},\mathbf{1})_{-1}\oplus(\mathbf{1},\mathbf{3})_{-1}\,.
\end{equation}
This twist thus produces exactly one scalar supercharge, given by the $(\mathbf{1},\mathbf{1})_{0}$ above, which we call $Q$. From the point of view of $X_4$ this is precisely the Donaldson-Witten scalar supercharge discussed in Section~\ref{sec:4dN2}. From the point of view of $C$ this is the unique twist of the 2d $\cN=(0,2)$ supersymmetric theory \cite{Witten:1988xj}. In the twisted theory, the four-dimensional translations are $Q$-exact. The theory is therefore topological along $X_4$ and does not depend on its metric. On the other hand, the generator of anti-holomorphic translations on $C$, given by $\cP_{34}$, is $Q$-exact, while holomorphic translations are not. The theory is therefore holomorphic along $C$.

To see more clearly how a chiral algebra can arise from this twist, let us proceed in two steps. First, we consider the holomorphic twist along $C$, coupling the rotations on $C$ with $\U(1)_r$ symmetry. This twist was studied in different contexts in \cite{Witten:1997sc,Maldacena:2000mw,Gaiotto:2009we}. The quantum number related to the twisted rotation on $C$ is given by $h_1^\prime = h_1+r$. As is apparent from \eqref{twistrep}, the twisted theory has eight supercharges transforming as scalars along $C$. Using these supercharges together with the translations and rotations along $X_4$, we can form a 4d $\cN=2$ subalgebra of the 6d $\cN=(2,0)$ superconformal algebra. The Poincar\'e supercharges of this 4d $\cN=2$ subalgebra are given by the following subset of the 16 Poincar\'e of the 6d $\cN=(2,0)$ theory
\begin{align}\label{eq:4d6dQs}
\cQ^1_+ &= \cQ_{31} \,, & \cQ^1_- &= \cQ_{32} \,, & \cQ^2_+ &= \cQ_{41} \,, & \cQ^2_- &= \cQ_{42} \,, \\
\widetilde{\cQ}_{1\dot{+}} &= -\cQ_{23} \,, & \widetilde{\cQ}_{1\dot{-}} &= -\cQ_{24} \,, & \widetilde{\cQ}_{2\dot{+}} &= -\cQ_{13} \,, & \widetilde{\cQ}_{2\dot{-}} &= -\cQ_{14} \,.
\end{align}
The generators of the 4d $\cN=2$ $R$-symmetry are given by \eqref{Rid}. The conformal supercharges and the remaining generators of this subalgebra are summarized in Appendix~\ref{App:subalg}. The second step in the topological-holomorphic twist of interest is to perform the Donaldson-Witten topological twist on $X_4$. The result of this twist will be a chiral algebra on $C$. However, this algebra of local operators of the topological-holomorphic theory is not very interesting. As in the work of Kapustin \cite{Kapustin:2006hi}, one finds that the product of two local operators can not contain any singularities. This can be seen as follows. Since the theory is topological on $X_4$ the operators can be inserted at any point on the four-manifold and thus they cannot have a singular OPE on $C$ and therefore do not have an interesting chiral algebra.  This is a consequence of Hartog's extension theorem for functions of multiple complex variables. In order to obtain an interesting OPE that allows for singularities, we must modify the topological-holomorphic twist to eliminate the freedom to move freely on $X_4$ and fix the local operators to a point on $X_4$. This is precisely what the $\Omega$-deformation summarized in Section~\ref{sec:4dN2} achieves! Therefore by adding an $\Omega$-deformation to the topologically twisted theory on $X_4$ all local operators localize to the fixed points on $X_4$ allowing for a non-trivial chiral algebra along $C$. To ensure that this chiral algebra is the same as the one in \cite{Beem:2014kka} we need to take care of one final subtlety. In \cite{Beem:2014kka} the 6d $\mathcal{N}=(2,0)$ theory was not deformed, i.e. the chiral algebra arises as a subsector of the 6d SCFT on $\mathbf{R}^6$. Therefore we should make sure that when we apply the $\Omega$-deformed topological-holomorphic twist on $X_4\times C$ described above, we do not deform the path integral of the 6d $\mathcal{N}=(2,0)$ SCFT. This can be achieved by taking $C=\mathbf{R}^2$ and, as described in detail in Section~\ref{subsec:4dN2Omega}, $X_4=\mathbf{R}^4$ with an $\Omega$-deformation with $\epsilon_1 = \epsilon_2=\f{\epsilon}{2}$.

After we have specified in detail the procedure for performing the topological-holomorphic twist on $\mathbf{R}^4_{\epsilon,\epsilon} \times \mathbf{R}^2$ it is important to study the resulting invariant supercharge and its cohomology. Fortunately, we have already done the necessary calculations. We simply need to take the supercharge in \eqref{Qom} and use \eqref{eq:4d6dQs} and Appendix~\ref{app:20} to express it in terms of the supercharges in the 6d $\mathcal{N}=(2,0)$ superconformal algebra. We find that the invariant supercharge after the $\Omega$-deformed topological-holomorphic twist on $\mathbf{R}^4_{\epsilon,\epsilon} \times \mathbf{R}^2$ is
\begin{equation}\label{eq:Qomega6d}
Q_{\Omega} = \cQ_{13}-\cQ_{24}-\epsilon\left(\cS_3^4+\cS_4^3\right)\,,
\end{equation}
and therefore as emphasized above does not deform the SCFT. One can check explicitly that indeed this supercharge squares to a twisted rotation of the 6d $\mathcal{N}=(2,0)$ theory
\begin{equation}
-\f{1}{2\epsilon}Q_\Omega^2 = \left( {\cM_3}^3-{\cM_4}^4 +\cR_{14}-\cR_{23}\right) = \f12 (\cL_2-\cL_3) -2R\,.
\end{equation}

As discussed in Section~\ref{subsec:4dN2Omega}, the cohomology of $Q_\Omega$ can be equivalently characterized by the cohomology  $Q_{\Omega}^{(1)}$ or $Q_{\Omega}^{(2)}$, where $Q_{\Omega}=Q_{\Omega}^{(1)}+Q_{\Omega}^{(2)}$ and $\{Q_{\Omega}^{(1)},Q_{\Omega}^{(1)}\}=\{Q_{\Omega}^{(2)},Q_{\Omega}^{(2)}\} = 0$. Using \eqref{eq:Qomega6d} we find that these supercharges are given explicitly by 
\begin{equation}\label{Q12BRvR}
Q_{\Omega}^{(1)} = \cQ_{13}-\epsilon \cS_3^4\,,\qquad Q_{\Omega}^{(2)} = -\cQ_{24}-\epsilon \cS_4^3\,.
\end{equation}
As remarked in Footnote~\ref{ftn:SU2} above there is some freedom in how to embed the $\SU(2)_R$ symmetry of the 4d $\mathcal{N}=2$ superconformal algebra in the $\USp(4)$ R-symmetry of the 6d $\mathcal{N}=(2,0)$ SCFT. If we make the alternative $\SU(2)_R$ choice described in Footnote~\ref{ftn:SU2} we can repeat the same topological-holomorphic twist accompanied by an $\Omega$-deformation. This leads to an alternative $\Omega$-deformed supercharge given by
\begin{equation}\label{eq:tQomega6d}
\widetilde{Q}_{\Omega} = -\cQ_{14}-\cQ_{23}-\epsilon\left(\cS_3^3-\cS_4^4\right)\,.
\end{equation}
The cohomology of this supercharge can be studied in a similar way by decomposing it as $\widetilde{Q}_{\Omega} = Q_{\Omega}^{(3)}+Q_{\Omega}^{(4)}$ with 
\begin{equation}\label{Q34BRvR}
Q_{\Omega}^{(3)} = -\cQ_{14}-\epsilon \cS_3^3\,,\qquad Q_{\Omega}^{(4)} = -\cQ_{23}+\epsilon \cS_4^4\,.
\end{equation}
We now notice that the $\Omega$-deformed supercharges described above can be mapped to the four supercharges $\mathbbmtt{Q}_i$ used to define the cohomology leading to the chiral algebra in \cite{Beem:2014kka}. The precise map is given by
\begin{equation}
\mathbbmtt{Q}_1 = Q_{\Omega}^{(1)}\,, \qquad \mathbbmtt{Q}_2 = -Q_{\Omega}^{(3)}\,, \qquad \mathbbmtt{Q}_3 = -Q_{\Omega}^{(4)}\,, \qquad \mathbbmtt{Q}_4 = -Q_{\Omega}^{(2)}\,.
\end{equation}
Since the supercharges $\mathbbmtt{Q}_i$ are the starting point for the construction of the chiral algebra in \cite{Beem:2014kka} we have shown that our procedure to perform a topological-holomorphic twist accompanied by an $\Omega$-deformation will eventually lead to the same chiral algebra. To confirm this more explicitly one can show that the generators of the left-moving $\sl(2)$ algebra on $C$ are given by
\begin{equation}
L_0 = \f12(\cH + \cL_1)\,, \qquad  L_{+1} = \cK^{21}\,, \qquad  L_{-1} = \cP_{12}\,,
\end{equation}
and are closed under $Q_{\Omega}$ and $\widetilde{Q}_{\Omega}$. As in \cite{Beem:2014kka} one can also appropriately modify the right-moving $\sl(2)$ algebra on $C$ using $\SU(2)_R$ as
\begin{equation}
\widehat{L}_0 = \f12\left(\cH - \cL_1\right)+\f{\epsilon}{2} \left(\cR_{14}+\cR_{23}\right)\,, \qquad \widehat{L}_{+1} = \cK^{43}-\epsilon\,\cR_{12}\,, \qquad \widehat{L}_{-1} = \cP_{34} + \epsilon\, \cR_{34}\,.
\end{equation}
It is then easy to show that these generators are exact under $Q_\Omega$ and $\widetilde{Q}_\Omega$
\begin{equation}
\begin{split}
\widehat{L}_{0} &=\frac{1}{4}\{ Q_\Omega , Q_\Omega^{\dagger} \} = \frac{1}{4}\{ \widetilde{Q}_\Omega , \widetilde{Q}_\Omega^{\dagger} \}\,,\\
\widehat{L}_{-1} &=\{ Q_\Omega , \cQ_{33}  \} = \{ Q_\Omega , \cQ_{44}  \} =-\{ \widetilde{Q}_\Omega , \cQ_{34}  \} = \{ \widetilde{Q}_\Omega , \cQ_{43}  \}\,, \\
\widehat{L}_{+1} &=-\{ Q_\Omega , \cS_{2}^3  \} = -\{ Q_\Omega , \cS_{1}^4  \} =-\{ \widetilde{Q}_\Omega , \cS_{2}^4  \} = \{ \widetilde{Q}_\Omega , \cS_{1}^3  \}\,.
\end{split}
\end{equation}
These results are of course compatible with \cite{Beem:2014kka} and dictate how to construct correlation functions of local operators inserted in the chiral algebra plane $C$.
	
To find the quantum numbers of a local operator inserted at the origin, $\mathcal{O}(0)$, that belongs to the cohomology of the supercharges $Q_{\Omega}^{(i)}$, $i=1,2,3,4$, and thus to the chiral algebra, we need to impose the identities (no sum over repeated indices)
\begin{equation}
[\{Q_{\Omega}^{(i)},(Q_{\Omega}^{(i)})^{\dagger}\},\mathcal{O}(0)]=0\,, \qquad [\{Q_{\Omega}^{(i)},Q_{\Omega}^{(j)}\},\mathcal{O}(0)]=0\,, ~~ i\neq j\,.
\end{equation}
Evaluating these commutators explicitly one finds that the quantum numbers of $\mathcal{O}(0)$ have to obey
\begin{equation}\label{eq:BPSoperators}
\Delta = 2R + h_1\,, \qquad r=0\,, \qquad h_2=h_3\,.
\end{equation}
Operators with these quantum numbers belong to (semi-)short representations of the $\mathcal{N}=(2,0)$ superconformal algebra. As could be anticipated these are precisely the operators studied in \cite{Beem:2014kka} and summarized in their Table 1.

\subsection{Intermezzo: lifting the $\Omega$-deformation to 6d}
\label{subsec:Omega6d}

We used a sleight of hand in the discussion above which we will address here.\footnote{We are grateful to Chris Beem for an important conversation about this issue.} When we write down the $\Omega$-deformed supercharge in \eqref{eq:Qomega6d} we seem to treat the supercharges on the right-hand side as in Section~\ref{subsec:4dN2Omega}, i.e. as if arising from three-dimensional spatial integrals of supercurrents, see \eqref{eq:VGcontract4dN2}. This is certainly not appropriate for supercharges acting on the 6d $\mathcal{N}=(2,0)$ SCFT. An alternative way to state the subtlety is that after we perform the holomorphic twist on $C$ we take $C$ to be non-compact, i.e. $C=\mathbf{R}^2$. In this section, we show how to resolve this problem and lift the $\Omega$-deformation in Section~\ref{subsec:4dN2Omega} to six dimensions.	
	
As in Section~\ref{subsec:4dN2Omega} it is useful to define the (conformal) supercharges in terms of supercurrents as
\begin{equation}
\cQ_{Aa} = \int \dd^5 x \, \mathcal{G}_{Aa} \,,\qquad 
\cS_{A}^{a} = \int \dd^5 x \,  \left(\overline{\Sigma}_\mu\right)^{ab}x^\mu \mathcal{G}_{Ab}\,.	
\end{equation}
The 4d scalar and vector supercharges of the Donaldson-Witten used in \eqref{Qom} and \eqref{eq:GDWdef} can be easily written in terms of the 6d supercharges as\footnote{In this subsection we sometimes use the superscript $^{(6d)}$ to emphasize that a given supercharge has been properly obtained by a 5d spatial integral of a supercurrent.}
\begin{equation}
	\begin{split}
		Q^{(6d)} &= \cQ_{13}-\cQ_{24} \,,\\
		G^{(6d)}_{X_4} &= i\bar{\Sigma}_{m}^{ab}\cQ_{Ab}M^A_{a}\dd x^m \,.
	\end{split}
\end{equation}	
Here $m=1,\dots,4$, the $\Sigma$ matrices are given in Appendix~\ref{App:bispin} and the matrix $M$ is defined as
\begin{equation}
M = \begin{pmatrix}
0&0&0&0\\0&0&0&0\\0&0&0&1\\0&0&-1&0
\end{pmatrix}\,.
\end{equation}
This matrix couples the $\SU(4)$ indices to the $\USp(4)$ indices and implements the topological twist along $X_4$. It can be checked explicitly that, with respect to the twisted rotations, $Q^{(6d)}$ transforms as a scalar along $X_4$ and $C$, i.e. it belongs to the $\left(\mathbf{1},\mathbf{1 }\right)_{0}$ representation in \eqref{twistrep}. Similarly $G^{(6d)}_{X_4}$ transforms as a vector along $X_4$ and a scalar along $C$ and thus corresponds to the $\left(\mathbf{2},\mathbf{2}\right)_{0}$ representation in \eqref{twistrep}.

The holomorphic twist on $C$ leads to another vector supercharge in the $\left(\mathbf{1},\mathbf{1}\right)_{-1}$ representation in \eqref{twistrep} which is given by
\begin{equation}
G^{(6d)}_{C} = i\bar{\Sigma}_{p}^{ab}\cQ_{Ab}N^A_{a}\dd x^p = \left(\cQ_{33}-\cQ_{44}\right)\dd z\,,
\end{equation}
where $p=5,6$, $z=x_5+i x_6$, and the matrix $N$ is defined as
\begin{equation}
N = \begin{pmatrix}
0&0&0&0\\0&0&0&0\\0&0&0&1\\0&0&1&0
\end{pmatrix}\,.
\end{equation}
To obtain the $\Omega$-deformed supercharge we now proceed as in Section~\ref{subsec:4dN2Omega}, namely we need to contract the vector supercharge with an appropriate Killing vector. To do this properly in 6d we need to use the vector supercharge $G^{(6d)}=G^{(6d)}_{X_4} + G^{(6d)}_C$. For the Killing vector on $\mathbf{R}^6$ we choose
\begin{equation}\label{eq:Vvec6d}
V^\mu = \Omega^{\mu\nu}x_\nu\,,
\end{equation}
where the matrix $\Omega$ is defined as
\begin{equation}\label{eq:omega6dmatrix}
\Omega = \begin{pmatrix}
0&\epsilon&0&0&0&0\\-\epsilon&0&0&0&0&0\\0&0&0&\epsilon&0&0\\0&0&-\epsilon&0&0&0\\0&0&0&0&0&\epsilon\\0&0&0&0&-\epsilon&0
\end{pmatrix}\,.
\end{equation}
The vector $V$ thus parameterizes rotations in the three orthogonal planes $\mathbf{R}^2\times \mathbf{R}^2\times\mathbf{R}^2 \subset X_4\times C$. On $X_4$ we recognize the familiar $\Omega$-deformation from Section~\ref{subsec:4dN2Omega} which is now accompanied by an additional $\Omega$-deformation on $C$. With this at hand we can then write the 6d $\Omega$-deformed supercharge $Q_{\Omega}^{(6d)}$ as 
\begin{equation}\label{eq:QOmega6dpr}
	\begin{split}
		Q_{\Omega}^{(6d)} &= Q^{(6d)}+ \iota_V G^{(6d)}\\
		&=  Q^{(6d)}+\int\dd^5 x\, \left(i\Omega^{mn}x_n\left(\bar{\Sigma}_m \right)^{ab} \mathcal{G}_{Ab}M^A_a + i\Omega^{pq}x_q\left(\bar{\Sigma}_p\right)^{ab} \mathcal{G}_{Ab}N^A_a\right)\\ 
		&=\cQ_{13}-\cQ_{24}-\epsilon(\cS_{3}^4+\cS_4^3)\,.
	\end{split}
\end{equation}
Proceeding in a similar fashion we can find another $\Omega$-deformed supercharge $\widetilde{Q}_{\Omega}^{(6d)}$ given by
\begin{equation}\label{eq:tQOmega6dpr}
	\begin{split}
		\widetilde{Q}_{\Omega}^{(6d)} &=  Q^{(6d)}+\int\dd^5 x\, \left(i\Omega^{mn}x_n\left(\bar{\Sigma}_m \right)^{ab} \mathcal{G}_{Ab}\widetilde{M}^A_a + i\Omega^{pq}x_q\left(\bar{\Sigma}_p\right)^{ab} \mathcal{G}_{Ab}\widetilde{N}^A_a\right)\\ 
		&=-\cQ_{14}-\cQ_{23}-\epsilon\left(\cS_3^3-\cS_4^4\right)\,.
	\end{split}
\end{equation}
where $\Omega$ is the same matrix as in \eqref{eq:omega6dmatrix} while $\widetilde{M}$ and $\widetilde{N}$ are given by 
\begin{equation}
\widetilde{M} = \begin{pmatrix}
0&0&0&0\\0&0&0&0\\0&0&-1&0\\0&0&0&-1
\end{pmatrix}\,, \qquad
\widetilde{N} = \begin{pmatrix}
0&0&0&0\\0&0&0&0\\0&0&-1&0\\0&0&0&1
\end{pmatrix}\,.
\end{equation}
We have thus arrived at the same $\Omega$-deformed supercharges as the one in \eqref{eq:Qomega6d} and \eqref{eq:tQomega6d} but with the benefit of a more rigorous 6d derivation. We thus conclude that the chiral algebra of \cite{Beem:2014kka} can be obtained from the cohomology of the supercharge of the topological-holomorphic twist of the 6d $\cN=(2,0)$ SCFT on the $\Omega$-deformed $\mathbf{R}^6_{\epsilon_1,\epsilon_2,\epsilon_3}$ with $\epsilon_1=\epsilon_2=\epsilon_3$.
	
A similar subtlety arises in the procedure of \cite{Oh:2019bgz,Jeong:2019pzg} to obtain the chiral algebra associated with every 4d $\mathcal{N}=2$ SCFT by employing an $\Omega$ deformation. We show how to address this subtlety in Appendix \ref{App:OhYagi} .

\subsection{Comments on supersymmetric localization}
\label{subsec:susyloc}

So far we have presented a procedure which involves topological twists and an $\Omega$-deformation of the 6d $\mathcal{N}=(2,0)$ SCFT and leads to a set of invariant supercharges which are identical to the ones used in \cite{Beem:2014kka} to arrive at a chiral algebra on $C$. This can be viewed as the kinematical underpinning for the derivation of this chiral algebra. To answer the dynamical question of the precise form of the chiral algebra one may adopt the approach followed in \cite{Beem:2014kka} which, in short, amounts to an educated guess supported by highly non-trivial consistency checks. Our construction above offers, at least in principle, an alternative route to the derivation of the chiral algebra. This could proceed as follows. We start with the path integral of the 6d $\mathcal{N}=(2,0)$ SCFT and implement the $\Omega$-deformation described in the previous section. This selects the supercharge $Q_{\Omega}$ in \eqref{eq:QOmega6dpr} which we could then use to perform equivariant localization along the lines of \cite{Nekrasov:2002qd,Nekrasov:2003rj}. It is natural to expect that the full path integral will then localize to a theory defined on $C$ spanned by the BPS operators in the $\mathcal{N}=(2,0)$ SCFT with quantum numbers obeying \eqref{eq:BPSoperators}. This localized path integral would then serve as a generating functional for the correlation functions of the chiral algebra in \cite{Beem:2014kka}. Unfortunately, this supersymmetric localization calculation is hard to perform explicitly. To understand why, recall that the general 6d $\mathcal{N}=(2,0)$ SCFT, labeled by a choice of a simply laced Lie algebra $\mathfrak{g}=\{A_N,D_N,E_{6,7,8}\}$, is intrinsically strongly coupled and lacks a (known) Lagrangian formulation that manifests all its symmetries. This prohibits a direct derivation by path integral methods of the conjecture in \cite{Beem:2014kka} that the chiral algebra associated to this 6d $\mathcal{N}=(2,0)$ SCFT is the $\mathcal{W}_{\mathfrak{g}}$ algebra. Nevertheless, it is possible to find supporting evidence that this path integral localization procedure will lead to the expected answer. As we show in Section~\ref{subsec:6danomint}, directly following \cite{Alday:2009qq}, upon an equivariant integration of the anomaly polynomial of the 6d $\mathcal{N}=(2,0)$ SCFT on $\mathbf{R}^4_{\epsilon_1,\epsilon_2}\times \mathbf{R}$ for a general choice of $\mathfrak{g}$ and after setting $\epsilon_1=\epsilon_2$ one indeed finds the central charge of a $\mathcal{W}_{\mathfrak{g}}$ algebra.
	
The free tensor multiplet lies in an ultra-short representation which contains five real scalars $\Phi_I$, two Weyl fermions $\lambda_{aA}$, and a two-form with self-dual field strength $w_{[ab]}^+$. The scalars transform in the $\mathbf{5}$ of $\so(5)$, which satisfy $\square \Phi^I=0$ and have $\Delta=2$. The two Weyl fermions transform in the $(\mathbf{4},\mathbf{4})$ of $\so(6)\times \so(5)$ and are subject to a symplectic Weyl reality condition. They satisfy the free Dirac equation and have scaling dimension $\Delta=\f52$. Finally, the two-form $w_{[ab]}^+$ has scaling dimension $\Delta=3$ and self-dual field strength $H=\star H=\dd w$. The field strength is both closed and co-closed, $\dd H = \dd\star H = 0$. In \cite{Beem:2014kka} it was shown that there is only one operator in the chiral algebra of the free tensor multiplet, namely the $\so(5)_R$ highest weight state of the five scalar fields, and its OPE forms an affine $\U(1)$ algebra. Here we outline how the same result should arise using supersymmetric localization in the $\Omega$-background described above. 
	
Due to the self-duality of the field strength, it is not possible to write down a Lagrangian for the free tensor multiplet. However, one can write down a free Lagrangian complemented with a self-duality condition $H^-=0$, see for example \cite{Arvidsson:2003ya}. The bosonic part of this Lagrangian reads
\begin{equation}
S = \int \left[ -\f12 \dd \Phi^I \wedge \star\dd \Phi^I - 2 H\wedge H + \f{i}{2}\star\left(\lambda^a\slashed{\partial}\lambda_a \right)\right]\,.
\end{equation}
To obtain a chiral algebra we put this theory on a manifold of the form $X_4\times C$. On this space, we can consider the manifestly self-dual reduction ansatz for the three-form field strength
\begin{equation}
H = F \wedge \dd x + \star_4 F \wedge \dd y + \star_4 \dd\Phi^6 + \dd\Phi^6\wedge \dd x\wedge \dd y\,, 
\end{equation}
where $z=x+iy$ is a complex coordinate on $C$, $\Phi^6$ is a scalar, $F$ is a two-form on $X_4$ and $\star_4$ is the four-dimensional Hodge star operator. With this reduction ansatz we see that the equation of motion for the three-form give rise to the Maxwell equation in four dimensions\footnote{The theory has become effectively four-dimensional due to the additional $\Omega$-deformation on $C$ discussed in Section~\ref{subsec:Omega6d}.}
\begin{equation}
\dd H = \dd\star H = 0 \quad \Rightarrow \dd F = \dd \star F = 0\,,
\end{equation}
and the Klein-Gordon equation for a massless scalar $\Phi_6$. Together with the remaining scalars and fermions, this results in the field content of $\cN=4$ $\U(1)$ SYM defined on $X_4$. We are thus led to conclude that supersymmetric localization of the $\cN=(2,0)$ free tensor multiplet on $\mathbf{R}^{4}_{\epsilon,\epsilon} \times C$ with a topological-holomorphic twist should be equivalent to supersymmetric localization $\cN=4$ $\U(1)$ SYM $\mathbf{R}^{4}_{\epsilon,\epsilon}$ with the Donaldson-Witten twist. To arrive at the final result we can appeal to the results in \cite{Dijkgraaf:2007sw} where it was argued that the theory resulting from this supersymmetric localization is the same as that of a chiral fermion which can be thought of as moving on the chiral algebra plane. While it will be nice to derive this result more rigorously, we will take it at face value and proceed to bosonize this chiral fermion and thus find a $\uu(1)$ Kac-Moody algebra on $C$. This is in harmony with the result in \cite{Beem:2014kka} for the chiral algebra arising from the $\cN=(2,0)$ free tensor multiplet. 

Clearly, it is desirable to perform the supersymmetric localization with more rigor for the general case of a 6d $\mathcal{N}=(2,0)$ SCFT of type $\mathfrak{g}$. Perhaps the approach outlined in \cite{Cordova:2016cmu} to use 5d $\mathcal{N}=2$ SYM obtained by a reduction of the 6d $\mathcal{N}=(2,0)$ SCFT will be fruitful in this regard. It is clear that there must be a relation between the $\mathcal{W}_{\mathfrak{g}}$ algebra appearing in the AGT correspondence and the chiral algebra of the same type arising in the construction of \cite{Beem:2014kka}. We hope that the procedure outlined above will shed more light on this connection.
	
\subsection{Orbifolds and defects}
\label{subsec:deforb}

One way to generalize the construction above is to include orbifolds acting on the space $X_4$. The orbifold action should be compatible with the Killing vector in \eqref{eq:Vvec6d} used
to define the $\Omega$-deformation. This implies that if we think of $X_4$ as a two-dimensional complex space $\mathbf{C}^2$ with coordinates $(z,w)$ then we should consider $\mathbf{Z}_k$ orbifolds defined by the action $(z,w) \rightarrow (\e^{2\pi i/k}z,\e^{-2\pi i/k}w)$. Similarly, it is possible to generalize the four-dimensional construction of \cite{Oh:2019bgz,Jeong:2019pzg} to include an Abelian orbifold of the plane transverse to the chiral algebra plane, i.e. take $\Sigma= \mathbf{R}^2/\mathbf{Z}_k$ as the surface transverse to $C$. As discussed in detail above, in the absence of orbifolds the 6d $\cN=(2,0)$ SCFT is not deformed by the topological-holomorphic twist and the $\Omega$-deformation. The same is true for the construction of \cite{Oh:2019bgz,Jeong:2019pzg}. This is crucial to ensure that the resulting chiral algebra is indeed a property of a given 6d $\cN=(2,0)$ or 4d $\cN = 2$ SCFT. In the presence of the $\mathbf{Z}_k$ orbifold, this is no longer true and thus the resulting chiral algebras should be associated with an appropriate deformation of the 6d $\cN=(2,0)$ or 4d $\cN = 2$ SCFT. 

It is tempting to speculate that for the 6d $\cN=(2,0)$ theory on this orbifold space it is possible to use supersymmetric localization to derive the corresponding chiral algebra.
The results of \cite{Kimura:2011zf,Bonelli:2012ny} may be useful in this regard. In the absence of such an explicit calculation we can try to formulate an educated guess for the chiral algebras arising
from this orbifold construction. To formulate the conjecture we draw inspiration from the results in \cite{Belavin:2011pp,Nishioka:2011jk,Bonelli:2011jx} on the generalizations of the AGT correspondence to theories on 4d orbifolds. The natural conjecture for the chiral algebra obtained from localizing the 6d $\cN=(2,0)$ theory of type $\mathfrak{g}=A_{N-1}$ on the $\Omega$-deformed $\mathbf{R}^4/\mathbf{Z}_k \times C$ is that it should be the same as the algebra of a system of generalized para-fermions known as ${\rm RCFT}[A_{k-1},A_{N-1}]\oplus {\rm RCFT}[A_{N-1},A_{k-1}]$ \cite{Cecotti:2010fi}, see \cite{Christe:1988vc,Bowcock:1988vs,Nam:1989ya,Nemeschansky:1991pr}. More generally for $\cN=(2,0)$ theories of type $\mathfrak{g}$ one should find a system of generalized para-fermions of type ${\rm RCFT}[A_{k-1},\mathfrak{g}]\oplus {\rm RCFT}[\mathfrak{g},A_{k-1}]$. Indeed, this appears to be in line with the discussion in \cite{Dijkgraaf:2007sw}, where the localization of the four-dimensional theory resulting from the abelian $\cN=(2,0)$ theory was studied. In Section~\ref{subsec:orbequiv} we discuss the results in \cite{Nishioka:2011jk} which  provide supporting evidence for this conjecture by computing the chiral algebra central charge via an explicit equivariant integration of the anomaly polynomial of the 6d $\cN=(2,0)$ SCFT of type $\mathfrak{g}$.

Similarly one should be able to construct a chiral algebra starting from a 4d $\cN = 2$ theory on $\mathbf{R}^2/\mathbf{Z}_k \times \mathbf{R}^2$ along the lines of \cite{Oh:2019bgz,Jeong:2019pzg}. In this case, however, we are not able to formulate a guess for the resulting chiral algebra. Nevertheless, in Section~\ref{subsec:orbequiv} we calculate the central charge of this chiral algebra using equivariant integration of the 4d $\cN = 2$ anomaly polynomial. It would certainly be interesting to explore this in more detail.

Another way to enrich the story above is to include defects in the six-dimensional theory. In order to be compatible with the $\Omega$-deformation and topological-holomorphic twist of interest, one has to restrict to supersymmetric co-dimension two or co-dimension four defects. The co-dimension two defects admit an $\su(4^\star|2)$ superconformal algebra on their worldvolume and thus come equipped with their own chiral algebra. If the defect extends along $C$ and spans two directions in $X_4$ then it modifies the chiral algebra of the $\cN=(2,0)$ SCFT. Defects extending along all four directions of $X_4$ are interpreted as chiral vertex operators in the chiral algebra on $C$. The co-dimension four supersymmetric defects compatible with the chiral algebra should have $\su(2|2)$ worldvolume superconformal symmetry. Some properties of these superconformal defects have been studied in \cite{Beem:2014kka,Bullimore:2014upa} but there is more to be understood. This is especially true in the context of relating the chiral algebra in the presence of these defects with the role that similar defects play in the AGT correspondence, see \cite{LeFloch:2020uop} for a recent review. We hope that the $\Omega$-deformed topological-holomorphic twist discussed above will shed some new light on this.

\section{Central charges from the anomaly polynomial}
\label{sec:centralc}
	
In this final section, we show how one can extract the central charges of the chiral algebras arising as subsectors in 4d $\cN=2$ and 6d $\cN=(2,0)$ SCFTs. The calculation proceeds along the lines of \cite{Alday:2009qq} and amounts to integrating the anomaly polynomial of the SCFT over the $\Omega$-deformed space transverse to the chiral algebra plane.

\subsection{Chiral algebras in 4d $\cN=2$ SCFTs}
	
The anomaly polynomial of a 4d $\cN=2$ SCFT with conformal anomalies $a$ and $c$ and flavor anomaly $k_G$ is given by (see for example \cite{Bah:2018gwc}) 
\begin{equation}\label{eq:I6}
	\mathcal{I}_6 = (a-c)\left[ 8 c_1(F_r)^3- 2 c_1(F_r)p_1(T_4)\right] - 8 (2a-c)c_1(F_r)c_2(F_R)+2k_G c_1(F_r) c_2(F_G)\,.
\end{equation}
Here $c_{1,2}$ are the first and second Chern classes, $p_1$ is the Pontryagin class, and $F_r$, $F_R$ and $F_G$ are the field strengths associated to the background vector fields of the $\U(1)_r$, $\SU(2)_R$ and flavor symmetries, respectively. 
	
We consider a 4d $\cN=2$ SCFT on a manifold of the form $X_4 = \Sigma\times C$ and denote with $\lambda$ and $t$ the Chern roots of the tangent bundles of $\Sigma$ and $C$, respectively. Following the construction of \cite{Oh:2019bgz,Jeong:2019pzg}, we should perform the topological-holomorphic twist of Kapustin accompanied by an $\Omega$-deformation along $\Sigma$. To obtain the central charges we proceed in steps. First, we twist the theory along $\Sigma$ by coupling the $\U(1)_r$ background vector field to the spin connection. This is implemented by identifying the first Chern class of the R-symmetry bundle of the four-dimensional theory with the Chern root of the tangent bundle of $\Sigma$, 
\begin{equation}
c_1\left(F_r\right) = - \f12 \lambda\,,
\end{equation}
while the $\SU(2)_R$ bundle remains untouched and becomes the R-symmetry of the resulting 2d $\cN=(0,4)$ theory on $C$. We can then evaluate the anomaly polynomial for the twisted theory and integrate $t_\Sigma$ over $\Sigma$. This operation results in the following anomaly four-form,
\begin{equation}\label{eq:anom4form}
I_4 = - 2(2a-c) c_1(F_R)^2 \int_\Sigma \lambda + (a-c) p_1(TC)\int_\Sigma \lambda - k_G c_1(F_G)^2\int_\Sigma \lambda \,. 
\end{equation}
For a 2d $\cN=(0,4)$ theory with left- and right-moving central charges $c_L$ and $c_R$ and flavor symmetry at level $k_{2d}$, the anomaly polynomial takes the form (see for example \cite{Bobev:2017uzs})
\begin{equation}\label{anomd2}
I_4^{(0,4)} = \f{c_R}{6} c_1(F_R)^2 + \f{c_L-c_R}{24} p_1(TC) + 2 k_{2d} c_1(F_G)^2\,.
\end{equation} 
Comparing \eqref{eq:anom4form} and \eqref{anomd2} we find the following central charges for the theory on $C$ \footnote{In our conventions the supercurrents are right-moving and therefore $c_R$ is associated with the supersymmetric part of the theory.}
\begin{equation}\label{eq:2d04anom}
c_R = -12\,\chi(\Sigma) (2a-c)\,,\qquad c_L = -12\,\chi(\Sigma)c\,, \qquad k_{2d} = -\frac{\chi(\Sigma)}{2}k_G\,.
\end{equation}
Here $\chi(\Sigma) =  \int_\Sigma \lambda$ is the Euler characteristic of $\Sigma$.
	
This result holds for all four-manifolds of the form $\Sigma\times C$ where $\Sigma$ and $C$ are two-manifolds. To study the chiral algebra of interest, we want to consider a 4d $\mathcal{N}=2$ SCFT on $\mathbf{R}^2_{\epsilon} \times \mathbf{R}^2$ and perform the twist along $\Sigma = \mathbf{R}^2_\epsilon$ where $\mathbf{R}^2_\epsilon$ is the two-dimensional $\Omega$-background. To properly compute the Euler characteristic of the $\Omega$-background, we should proceed with care and consider the characteristic classes in the equivariant sense \cite{Bonelli:2009zp,Alday:2009qq,Bobev:2015kza}. We consider the action of a $\U(1)$ rotating the plane around the origin and call the equivariant parameter $\epsilon$. We can now use the Duistermaat-Heckman (or more general Berline-Vergne/Atiyah-Bott) localization formula
\begin{equation}
\int_M \alpha = \sum_p \frac{\alpha|_p}{e(N_p)}\,,
\end{equation}
where $\alpha$ is an equivariantly closed differential form. The sum runs over all fixed points of the torus action (in our case simply the $\U(1)$ action) and $e(N_p)$ is the equivariant Euler class of $N_p$, the normal bundle of $p$ in $\Sigma$. In our case, the only fixed point of the $\U(1)$ action is the origin and therefore we find the following equivariant Euler characteristic
\begin{equation}
	\chi(\mathbf{R}^2_\epsilon) = \int_{\mathbf{R}^2_\epsilon} \lambda = 1\,.
\end{equation}
Finally, as described in \cite{Oh:2019bgz,Jeong:2019pzg}, to obtain a chiral algebra on $C$ we have to perform a holomorphic twist of the 2d $\mathcal{N}=(0,4)$ theory with anomalies in \eqref{eq:2d04anom} with respect to the Cartan of the $\SU(2)_R$ symmetry. This makes the right-moving sector of the theory topological and thus we find 
\begin{equation}\label{4d2dcLR}
c_R = 0\,,\qquad c_L = -12 c\,, \qquad k_{2d} = -\frac{1}{2}k_G\,.
\end{equation}
These are precisely the chiral algebra anomalies derived by different methods in \cite{Beem:2013sza}. Note in particular that for a unitary 4d $\mathcal{N}=2$ SCFT we have $k_G>0$ and $c>0$ and thus we find a non-unitary chiral algebra.
	
\subsection{Chiral algebras in 6d $\cN=(2,0)$ SCFTs}
\label{subsec:6danomint}
	
In six dimensions we can proceed in the same spirit. The calculation was performed in detail in \cite{Alday:2009qq} and here we summarize their results with particular emphasis on the $\epsilon_1=\epsilon_2$ limit relevant for our discussion.  We consider the 6d $\mathcal{N}=(2,0)$ theory of type $\mathfrak{g}$ on a manifold of the form $X_4\times C$. Following the construction of the chiral algebra in Section~\ref{sec:6dchiral}, we consider a topological-holomorphic twist of the 6d theory accompanied by an $\Omega$-deformation.
	
The anomaly polynomial of the 6d $\cN=(2,0)$ theory of type $\mathfrak{g}$, where $\mathfrak{g} =\{A_N,D_N,E_{6,7,8}\}$, can be written as \cite{Harvey:1998bx,Intriligator:2000eq,Yi:2001bz}
\begin{equation}\label{I8G}
I_8[\mathfrak{g}] = r_\mathfrak{g} I[1] + \frac{d_\mathfrak{g} h_\mathfrak{g}}{24} p_2({\rm NW})\,,
\end{equation}
where $r_\mathfrak{g}$, $d_\mathfrak{g}$ and $h_\mathfrak{g}$ are the rank, dimension, and Coxeter number of the simply-laced Lie algebra $\mathfrak{g}$, see Table \ref{tab:rdh}, and $I_8[1]$ is the anomaly eight-form of one M5-brane \cite{Witten:1996hc}
\begin{equation}\label{I81}
I_8[1] = \f{1}{48}\left[ p_2({\rm NW}) - p_2({\rm TW}) + \f14\left( p_1({\rm TW})-p_1({\rm NW}) \right)^2 \right]\,,
\end{equation} 
where NW and TW stand for the normal and tangent bundle to the M5-brane worldvolume and $p_k$ is the $k^{\rm th}$ Pontryagin class. 

\begin{table}[!htb]
	\centering
	\begin{tabular}{c|ccc}
		$\mathfrak{g}$ & $r_\mathfrak{g}$ & $d_\mathfrak{g}$ & $h_\mathfrak{g}$\\
		\hline
		$\Tstrut A_{N-1}$ & $N-1$ & $N^2-1$ & $N$\\
		$D_N$ & $N$ & $2N^2-N$ & $2N-2$ \\
		$E_6$ & 6 & 78 & 12 \\
		$E_7$ & 7 & 133 & 18 \\
		$E_8$ & 8 & 248 & 30
	\end{tabular}
	\caption{Rank, dimension and Coxeter number of the Lie algebras of type $\mathfrak{g}$.}
	\label{tab:rdh}
\end{table}
	
We proceed by twisting the theory along $X_4$ as described in Section \ref{sec:6dchiral}. The supercharges preserved under this twist form a 2d $\cN=(0,2)$ superalgebra with $\U(1)_r$ symmetry. We can determine the anomaly polynomial of the resulting two-dimensional theory by integrating the eight-form anomaly polynomial \eqref{I8G} over $X_4$. We denote the Chern root of the tangent bundle of $C$ with $t_C$ and use $\lambda_1$ and $\lambda_2$ for the Chern roots of the tangent bundle of $X_4$. The Chern roots of the normal bundle are $n_1$ and $n_2$. The topological twist along $X_4$ introduces the identification
\begin{equation}
n_2 = \lambda_1 + \lambda_2\,,
\end{equation}
and we can identify the R-symmetry of the two-dimensional theory on $C$ as $2c_1(F_R)~\to~n_1$. Expanding the Pontryagin classes as
\begin{align}
p_1(\rm TW) &= p_1(TC) + p_1(TX_4) \,, & 	p_2(\rm TW) &= p_1(TC)p_1(TX_4)\,, \\
p_1(\rm NW) &= 4c_1(F_R)^2 + p_1(TX_4) + 2e(X_4) \,, & 	p_2(\rm NW) &= 4 c_1(F_R)^2 (p_1(TX_4)+e(X_4))\,, \nonumber
\end{align}
where $e(X_4)$ is the Euler class of $X_4$, we can integrate the anomaly polynomial over $X_4$ to find the four-form anomaly polynomial
\begin{equation}
\begin{aligned}
I_4 &= \left[ \f{r_\mathfrak{g}+2d_\mathfrak{g} h_\mathfrak{g}}{12}\int_{X_4}p_1(TX_4) + \f{3r_\mathfrak{g}+4d_\mathfrak{g} h_\mathfrak{g}}{12}\int_{X_4}e(X_4) \right]c_1(F_R)^2 \\
&\hspace{3cm} -\f{r_\mathfrak{g}}{48}\left[ \int_{X_4}\left(p_1(TX_4)+e(X_4)\right)\right]p_1(TC)\,.
\end{aligned}
\end{equation}
Comparing this with \eqref{anomd2} we find the left- and right-moving central charge\footnote{The expression in  \eqref{anomd2} is also the anomaly polynomial for 2d $\mathcal{N}=(0,2)$ theories.}
\begin{equation}\label{6d2danomssdf}
\begin{aligned}
c_L &= \chi(X_4)r_\mathfrak{g} + \left( P_1(X_4)+2\chi(X_4) \right)d_\mathfrak{g} h_\mathfrak{g}\,,\\
c_R &= \f12\left( P_1(X_4) + 3\chi(X_4) \right)r_\mathfrak{g} + \left( P_1(X_4) + 2\chi(X_4) \right)d_\mathfrak{g} h_\mathfrak{g}\,,
\end{aligned}
\end{equation}
where $\chi(X_4) = \int_{X_4} e(TX_4)$ is the Euler number of $X_4$ and $P_1(X_4) = \int_{X_4} p_1(TX_4)$ is the integrated first Pontryagin class which is equal to three times the signature of $X_4$. 
	
So far we treated $X_4$ as a general four-manifold. Our main interest here is to study the theory on the $\Omega$-background, $X_4 = \mathbf{R}^4_{\epsilon_{1,2}}$. We take the action of the torus $\U(1)^2$ to rotate the two orthogonal planes in $\mathbf{R}^4$ with equivariant parameters $\epsilon_1$ and $\epsilon_2$. To compute the Euler number and integrated Pontryagin class we need to employ equivariant integration. Since the Chern classes of the two planes are given by $\epsilon_{1,2}$ we have $p_1(T\mathbf{R}^4_{\epsilon_{1},\epsilon_{2}}) = \epsilon_1^2+\epsilon_2^2$ and $e(T\mathbf{R}_{\epsilon_{1},\epsilon_{2}}^4)=\epsilon_1 \epsilon_2$. The only fixed point of the torus action is again the origin so we find
\begin{equation}
	P_1(\mathbf{R}^4_{\epsilon_{1},\epsilon_{2}}) = \frac{\epsilon_1^2 + \epsilon_2^2}{\epsilon_1\epsilon_2}\,,\qquad \chi(\mathbf{R}^4_{\epsilon_{1},\epsilon_{2}}) = 1\,.
\end{equation}
Substituting this result in \eqref{6d2danomssdf} we arrive at, see \cite{Alday:2009qq},
\begin{equation}\label{6dcLR}
\begin{aligned}
c_L &= r_\mathfrak{g}+(b+b^{-1})^2 h_\mathfrak{g} d_\mathfrak{g}\,, \\
c_R &= \frac{1}{2}((b+b^{-1})^2+1)r_\mathfrak{g}+(b+b^{-1})^2 h_\mathfrak{g} d_\mathfrak{g}\,,
\end{aligned}
\end{equation}
where $b^2 = \epsilon_1/\epsilon_2$. To complete the chiral algebra construction we need to implement a further holomorphic twist of the two-dimensional theory on $C$. This sets $c_R=0$ and leaves only a left-moving chiral algebra with central charge $c_L$. As emphasized in Section~\ref{sec:6dchiral} the chiral algebra of interest arises when we set $\epsilon_1=\epsilon_2$ or $b=1$. Setting $b=1$ in \eqref{6dcLR} one finds that $c_L=4 d_\mathfrak{g}h_\mathfrak{g}+r_\mathfrak{g}$ which is the expected central charge of the $\mathcal{W}_\mathfrak{g}$ algebra \cite{Beem:2014kka}.	
	
\subsection{Including orbifolds}
\label{subsec:orbequiv}
	
As discussed in Section~\ref{subsec:deforb}, we can generalize the setup of \cite{Beem:2013sza,Beem:2014kka} and replace $\Sigma = \mathbf{R}^2$ and $X_4 = \mathbf{R}^4$ by the orbifold spaces $\mathbf{C}/\mathbf{Z}_k$ and $\mathbf{C}^2/\mathbf{Z}_k$ where the orbifold action is $z \mapsto \e^{2\pi i / k} z$ and $(z,w)\mapsto (\e^{2\pi i / k} z,\e^{-2\pi i / k} w)$, respectively. To find the central charge of the corresponding chiral algebra we need to slightly modify the computations above when it comes to the evaluation of the equivariant Euler number and integrated first Pontryagin class.

In four dimensions we must compute the equivariant Euler characteristic on $\mathbf{R}^2_\epsilon/\mathbf{Z}_k$. We do this by passing to the $k$-fold cover of the orbifold. On this cover, the $\U(1)$ action has $k$ fixed points, one corresponding to each sheet. From the localization formula it then straightforwardly follows that the Euler characteristic is given by
\begin{equation}
\chi(\mathbf{R}^2_\epsilon/\mathbf{Z}_k) = \sum_i 1 = k\,,
\end{equation}
where $i$ runs over the fixed points. Inserting this result in the formulae for the central charges, \eqref{eq:2d04anom}, we find the central charge and current level of a chiral algebra obtained from an $\cN=2$ SCFT as
\begin{equation}
c_{2d} = -12 k c\,, \qquad\qquad k_{2d} = -\frac{k}{2} k_G\,.
\end{equation}
For $k=1$ we recover the result in \eqref{4d2dcLR}. It will be interesting to understand whether this two-dimensional chiral algebra has any relation to the one studied in Section 5 of \cite{Cecotti:2015lab}.

In the six-dimensional case, we proceed analogously. This calculation was in fact performed in \cite{Nishioka:2011jk} and here we briefly summarize their results. To compute the equivariant characteristic classes for $\mathbf{R}^4_{\epsilon_{1},\epsilon_{2}}/\mathbf{Z}_k$ it is more convenient to work with the resolved ALE space obtained by blowing up the singularity at the origin. The blowup of $\mathbf{C}^2/\mathbf{Z}_k$ is given as the standard blowup of an $A_{k-1}$ singularity, where we replace the singular point at the origin by a collection of $k-1$ $\mathbf{CP}^1$'s which are either completely disjoint or intersect in exactly one point. The intersection numbers are related to the Cartan matrix of the $A_{k-1}$ algebra. In the resolved space, the fixed points of the $\U(1)^2$ action can be mapped to the nodes of the Dynkin diagram of $A_{k-1}$. Therefore, we conclude that the blown-up geometry contains $k$ fixed points which contribute to the localization formula. It proves useful to introduce local coordinates for the $k$ fixed points given respectively by $(z_i , w_i) = (z^{k-i+1}w^{1-k} , z^{i-k}w^{k})$. If we parameterize the original $\U(1)^2$ action on $\mathbf{R}^4$ as $(z,w)\mapsto(\e^{\epsilon_{1}}z,\e^{\epsilon_{2}}w)$, then the torus action at the fixed points is given by $(z_i,w_i)\mapsto (\e^{\epsilon_1(i)}z_i,\e^{\epsilon_2(i)}w_i)$, where
\begin{equation}
\epsilon_1(i) = (k-i+1)\epsilon_1 + (1-i)\epsilon_{2}\,,\qquad \epsilon_2(i) = (i-k)\epsilon_1 + i \epsilon_{2}\,.
\end{equation}
Using these data, the equivariant Euler number and integrated Pontryagin class can be computed from the localization formula resulting in
\begin{align}
\chi(\mathbf{R}^4_{\epsilon_{1},\epsilon_{2}}/\mathbf{Z}_k) &= \sum_i 1 = k\,,\\ 
P_1(\mathbf{R}^4_{\epsilon_{1},\epsilon_{2}}/\mathbf{Z}_k) &= \sum_i \frac{\epsilon_1(i)^2 + \epsilon_2(i)^2}{\epsilon_1(i)\epsilon_2(i)} = \frac{1}{k}\frac{(\epsilon_1+\epsilon_2)^2}{\epsilon_1\epsilon_2}-2k\,.
\end{align}
Using this and \eqref{6d2danomssdf} we find the central charge
\begin{equation}\label{eq:paraWcc}
c_L = k \, r_{\mathfrak{g}} + \frac{d_{\mathfrak{g}} h_{\mathfrak{g}}}{k}(b+b^{-1})^2 \,.
\end{equation}
To obtain the chiral algebra central charge we need to set $b=1$. We then obtain a result for the central charge related to the central charge of the $k$-th para-$\mathcal{W}_{\mathfrak{g}}$ algebra as discussed in detail in \cite{Nishioka:2011jk}. For $k=1$ one recovers the result in \eqref{6dcLR}. For $k=2$ and $r_{\mathfrak{g}}=2$ one finds the central charge of $\mathcal{N}=1$ super Liouville theory which is in harmony with the results in \cite{Belavin:2011pp}. It will be very interesting to understand how to generalize the results of \cite{Beem:2014kka} and obtain a chiral algebra with the central charge in \eqref{eq:paraWcc} from the OPE of a suitable deformation of the 6d $\cN=(2,0)$ theory on $\mathbf{R}^4/{\mathbf{Z}_k} \times \mathbf{R}^2$.
		
\section{Discussion} 
\label{sec:discussion}

In this paper, we studied an alternative derivation of the chiral algebra associated with a 6d $\cN=(2,0)$ SCFT proposed in \cite{Beem:2014kka}. Our construction is similar in spirit to the one in \cite{Oh:2019bgz,Jeong:2019pzg} which is relevant for the chiral algebras associated with 4d $\cN=2$ SCFTs. An essential ingredient in our discussion is a topological-holomorphic twist of the 6d $\cN=(2,0)$ SCFT on the $\Omega$-deformed space $\mathbf{R}^4_{\epsilon_1,\epsilon_2}\times \mathbf{R}^2$. In the limit $\epsilon_1=\epsilon_2$ this operation does not deform the theory and the cohomology of the $Q_{\Omega}$ supercharge leads to the same chiral algebra on $ \mathbf{R}^2$ as in \cite{Beem:2014kka}. Furthermore, we also generalized this construction to include orbifolds in the space transverse to the chiral algebra plane and showed how to utilize equivariant integration of the anomaly polynomial of the 4d $\cN=2$ and 6d $\cN=(2,0)$ SCFTs to derive the chiral algebra central charges obtained by other means in \cite{Beem:2013sza} and \cite{Beem:2014kka}, respectively.

The most important open question is how to use our construction as a basis for a supersymmetric localization calculation in order to derive the proposal in \cite{Beem:2014kka} that the chiral algebra for the 6d $\cN=(2,0)$ theory of type $\mathfrak{g}$ is the $\mathcal{W}_g$ algebra. This result, and its orbifold generalization, are also supported by the central charge calculation in Section~\ref{subsec:6danomint}. As already discussed in Section~\ref{subsec:susyloc}, this direct derivation of the chiral algebra is difficult due to the absence of a Lagrangian of the 6d $\cN=(2,0)$ interacting SCFTs and it will be most interesting to circumvent this difficulty. One possible line of attack is to use the idea of \cite{Cordova:2016cmu}, see also \cite{vanLeuven:2017zfa}, and reduce the 6d $\cN=(2,0)$ theory to a 5d $\cN=2$ SYM theory in a non-trivial background. The knowledge of the explicit Lagrangian of the 5d SYM theory may then facilitate the supersymmetric localization calculation.

Many of the ingredients in our analysis are reminiscent of elements appearing in the AGT correspondence \cite{Alday:2009aq}. The twist we employ on $C$ is the same as the one in the class $\cS$ construction  \cite{Witten:1997sc,Gaiotto:2009we}, the difference being that in our setup $C$ is simply the Euclidean plane instead of a punctured Riemann surface. Moreover, it is well-known that the $\Omega$-deformation is intimately related to the Nekrasov partition function and therefore AGT \cite{Nekrasov:2010ka}. Finally, the appearance of the Liouville \cite{Alday:2009aq} and more general Toda \cite{Wyllard:2009hg} CFTs in the AGT setup is of course reminiscent of the $\cW$ algebras arising as chiral algebras in the construction of \cite{Beem:2014kka}. All of this suggests that it might be possible to generalize the results presented here to more general surfaces $C$ and combine them with supersymmetric localization to arrive at a derivation of the AGT correspondence. This may offer an alternative to the derivation in \cite{Cordova:2016cmu} of the Toda CFT from the 6d $\cN=(2,0)$ SCFT in which the role of the $\Omega$-background is more manifest. Inspired by these observations it is tempting to speculate that the 3d-3d correspondence, see \cite{Dimofte:2014ija} for a review, can be somehow leveraged to uncover a subset of operators in the 6d $\cN=(2,0)$ SCFT which span a 3d complex Chern-Simons theory. Perhaps combining the ideas discussed in \cite{Cordova:2013cea} together with a topological twist accompanied by an $\Omega$-deformation may be a fruitful strategy to pursue this.

It is natural to wonder also whether the idea to use a topological-holomorphic twist to arrive at an interesting cohomological sector with a non-trivial OPE can be generalized to other setups. While the chiral algebras proposed in \cite{Beem:2013sza} and \cite{Beem:2014kka} require the 4d and 6d theories from which they originate to be conformal and highly supersymmetric one may hope that there are generalizations of these ideas to theories with less supersymmetry or with broken conformal invariance. For example the 1d topological theory studied in \cite{Chester:2014mea} arising from 3d $\cN=4$ SCFTs can be extended to non-conformal theories \cite{Dedushenko:2016jxl}.\footnote{See also \cite{Panerai:2020boq}, where these results were extended to $\cN=4$ theories on more general 3d manifolds.} In addition, there are some hints that a version of the topological-holomorphic twist with interesting relations to chiral algebras arises also in 4d $\cN=1$ QFTs \cite{Johansen:1994aw,Johansen:1994ij,Johansen:1994ud}. It will be very interesting to understand whether using the $\Omega$-deformation accompanied by an appropriate topological twist may lead to new insights on these examples. Finally we should mention that a topological-holomorphic twist of the non-conformal 6d $\cN=(1,1)$ SYM on $X_4\times C$ with $X_4=\mathbf{R}^4_{\epsilon_1,\epsilon_2}$ was studied in \cite{Costello:2018txb}. This construction uses the Marcus, or geometric Langlands, twist of the effective 4d $\cN=4$ SYM theory on $X_4$  \cite{Marcus:1995mq,Kapustin:2006pk}. As explained in detail above, the construction employed in our work uses the Donaldson-Witten twist on $X_4$. It is then natural to ask whether there is any application of the Yamron-Vafa-Witten twist of $\cN=4$ in a similar context \cite{Yamron:1988qc,Vafa:1994tf}. We are not aware of an obvious candidate construction that results in such a topological-holomorphic twist and it will be interesting to understand the reason behind such a potential obstruction.

Another way to generalize our construction is to consider more general manifolds $X_4$ on which the topological twist accompanied by an $\Omega$-deformation can be performed. Indeed, for $X_4$ a general toric four-manifold one can perform the topological twist and define the $\Omega$-deformation using the $\U(1)^2$ isometry of the manifold. For compact $X_4$ one should find at low energies a 2d $\cN=(0,2)$ theory of the type studied in \cite{Benini:2013cda,Gadde:2013sca} and more recently in \cite{Hosseini:2020vgl}. For non-compact $X_4$ the $\Omega$-deformation may still give rise to new chiral algebra sectors on $C$ outside of the scope of \cite{Beem:2014kka}. However, for general $X_4$ it is a priori not clear if there exists a specific combination of equivariant parameters which leaves the 6d $\cN=(2,0)$ theory undeformed. Therefore in general these putative chiral algebras will not describe a protected subsector of the original 6d $\cN=(2,0)$ SCFT on $\mathbf{R}^6$.

The chiral algebras of \cite{Beem:2013sza} and \cite{Beem:2014kka} should have applications also in holography. This was explored initially in \cite{Bonetti:2016nma} and further in \cite{Costello:2018zrm} in the context of the 4d $\cN=2$ chiral algebras. In contrast, the 6d $\cN=(2,0)$ chiral algebras, and their orbifold generalizations discussed in our work, have not been studied holographically. It is clearly important to understand this better especially since the non-trivial OPE of the operators in the chiral algebra may lead to important insights into holography and perhaps the structure of M-theory \cite{Beem:2014kka,Beem:2015aoa,Chester:2018dga}. It will also be interesting to explore the relation between this and the appearance of the $\Omega$-deformation in AdS/CFT \cite{Bobev:2019ylk,Gaiotto:2019wcc,BenettiGenolini:2019wxg}.
	
\bigskip
\bigskip
\leftline{\bf Acknowledgments}
\smallskip
\noindent We have enjoyed useful discussions with Francesco Benini, Kiril Hristov, Leonardo Rastelli, and Junya Yagi. We are particularly grateful to Christopher John Beem for numerous informative conversations and patient explanations. We would also like to thank the anonymous referee for the careful reading of the paper and the useful feedback. The work of NB is supported in part by the Odysseus grant G0F9516N from the FWO. PB is supported by the STARS-StG grant THEsPIAN. FFG is a Postdoctoral Fellow of the Research Foundation - Flanders (FWO). This work is also supported in part by the KU Leuven C1 grant ZKD1118 C16/16/005.
	

\appendix
	
\section{Bi-spinors in four and six dimensions}
\label{App:bispin}
	
Here we collect some facts about the rotation groups in four and six dimensions and the bi-spinor notation used in the paper. The rotation groups in four and six dimensions enjoy the exceptional isomorphisms
\begin{equation}
\Spin(4) \simeq \SU(2)\times \SU(2)\,,\quad \text{and}\quad \Spin(6)\simeq \SU(4)\,.
\end{equation}
Thus there is a useful alternative way to represent various fields and operators of interest here. Instead of using the usual vector indices $\mu, \nu, ...$ we can us spinor indices transforming in the fundamental representations of $\SU(4)$ or $\SU(2)\times \SU(2)$. Furthermore, in the 6d $\cN=(2,0)$ theory the R-symmetry group is $\Spin(5)$ which also has a useful alternative representation due to the isomorphism
\begin{equation}
\Spin(5)\simeq \USp(4)\,.
\end{equation}
%
	
\subsection*{Bi-spinors in four dimensions}
	
In four dimensions, a Weyl spinor $\psi_\alpha$, $\alpha=1,2$ transforms in the fundamental representation of $\SL(2,\mathbf{C})$, the complex conjugate representation is denotes by $\psi_{\ald}$, $\ald = 1,2$. Both the dotted and undotted spinor indices are raised and lowered with the $\SL(2,\mathbf{C})$ invariant tensor $\epsilon_{\al\be}$ (and $\epsilon_{\ald\bed}$). In our conventions
	\begin{equation}
	\epsilon_{12} = \epsilon^{21} = \epsilon_{\dot{1}\dot{2}} = \epsilon^{\dot{2}\dot{1}} = 1\,.
	\end{equation} 
In this convention $\epsilon_{\al\be}\epsilon^{\be\gamma} = \delta_\al^\gamma$. The real forms of the complex Lie algebras $\sl(2,\mathbf{C})$ and $\su(2)_+\times \su(2)_-$ are the same. This allows us to use the same notation $(j_1,j_2)$ to label the representations, where $j_1$ and $j_2$ are the spins of the two $\su(2)$'s. In this way, the spinor representation and its complex conjugate representation correspond to $(\mathbf{2},\mathbf{1})$ and $(\mathbf{1},\mathbf{2})$. 
	
In Euclidean space, the spacetime rotation group is given by the orthonormal group $\SO(4)$ which has covering group $\Spin(4)$. This group in turn is isomorphic to $\SU(2)_+\times \SU(2)_-$. Exploiting this isomorphism we can rewrite vectors as bi-spinors
\begin{equation}
P_\mu (\sigma^\mu)_{\al\bed} = P_{\al\bed} = c_\alpha \tilde{c}_\bed\,.
\end{equation}
Here $c$ and $\tilde{c}$ are complex valued spinors transforming in the $(\mathbf{2},\mathbf{1})$ and $(\mathbf{1},\mathbf{2})$ representations of the Lorentz group. 
	
The Clebsch-Gordan coefficients $(\sigma^\mu)_{\al\ald}$ intertwining between the vector representation and the $(\mathbf{2},\mathbf{2})$ representation of $\SL(2,\mathbf{C})$ are given by
\begin{equation}\label{eq:defsigmaE}
(\sigma^\mu)_{\al\bed} = (\sigma^1,\sigma^2,\sigma^3,i{\bf 1}_{2\times 2})_{\al\bed} \,,
\end{equation}
where we have used the Pauli matrices
\begin{equation}\label{eq:Paulisigma}
\sigma^1 = \begin{pmatrix}
0&1\\1&0
\end{pmatrix}\,,\qquad\sigma^2 = \begin{pmatrix}
0&-i\\i&0
\end{pmatrix}\,,\qquad\sigma^3 = \begin{pmatrix}
1&0\\0&-1
\end{pmatrix}\,.
\end{equation}
The dual Clebsch-Gordan coefficients are defined as
\begin{equation}
(\bar{\sigma}^\mu)^{\ald\be} = (-\sigma^1,-\sigma^2,-\sigma^3,i{\bf 1}_{2\times 2})^{\ald\be}\,,
\end{equation}
and allow to transform a vector into a bi-spinor corresponding to the dual representation. The matrices $\sigma^\mu$ and $\bar{\sigma}^\mu$ satisfy the following relations: 
\begin{equation}
\begin{split}
		\bar{\sigma}^\mu\sigma^\nu + \bar{\sigma}^\nu\sigma^\mu &= -2 \eta^{\mu\nu}\,,\\
		(\bar{\sigma}^\mu)^{\ald\be}(\sigma_\mu)_{\gamma\dot{\delta}} &= - 2 \delta_\gamma^\be \delta_{\dot{\delta}}^{\ald}\,.
\end{split}
\end{equation}
To define the generators of $\Spin(4)$ rotations we first define the following matrices
\begin{equation}
{(\sigma^{\mu\nu})_\alpha}^\beta = \f14{(\sigma^\mu\bar{\sigma}^\nu -\sigma^\nu\bar{\sigma}^\mu)_\alpha}^\beta \,, \qquad {(\bar{\sigma}^{\mu\nu})^\bed}_\ald = \f14{(\bar{\sigma}^\mu\sigma^\nu -\bar{\sigma}^\nu \sigma^\mu)^\bed}_\ald\,.
\end{equation}
These matrices provide a spinorial representation of the Lorentz group, $M^{\mu\nu}=i\sigma^{\mu\nu}$ (or $M^{\mu\nu}=i\bar{\sigma}^{\mu\nu}$).
	
We can then decompose $M_{\mu\nu}$ into a self-dual and anti-self-dual part in the bi-spinor notation as
\begin{equation}
M_{\mu\nu} \rightarrow (\sigma^\mu)_{\al\ald}(\sigma^\mu)_{\be\bed}M_{\mu\nu} = {\cM}_{\al\ald,\be\bed} = \epsilon_{\al\be} \cM_{\ald\bed} + \epsilon_{\ald\bed} \cM_{\al\be}\,,
\end{equation}
so that $ {\cM_\ald}^\bed$ are the generators of $\SU(2)_-$ and ${\cM_\al}^\be$ the generators of $\SU(2)_+$. 
	
The bi-spinor notation allows to write the commutation relation for (super)conformal algebras of interest here in a unified manner irrespective of the signature of spacetime. The different choice of spacetime signature differ by the definition of the $\sigma$-matrices. In our analysis we mostly work in Euclidean signature and thus use the definition in \eqref{eq:defsigmaE}. To convert to Lorentzian mostly plus signature we should use the intertwining matrices:
\begin{equation}
(\sigma^\mu)_{\al\bed} = ({\bf 1}_{2\times 2},\sigma^1,\sigma^2,\sigma^3)_{\al\bed} \,,\qquad (\bar{\sigma}^\mu)^{\ald\be} = ({\bf 1}_{2\times 2},-\sigma^1,-\sigma^2,-\sigma^3)^{\ald\be}\,.
\end{equation}
%

\subsection*{Bi-spinors in six dimensions}
	
A similar construction can also be performed in six dimensions. The Lorentz group is $\Spin(6)\simeq \SU(4)$ and the six-dimensional spinors have four complex components, transforming in the fundamental of $\SU(4)$. We can rewrite the momentum vector as a bi-spinor
\begin{equation}
P_\mu (\Sigma^\mu)_{ab} = P_{ab} = c_a c_b\,,
\end{equation} 
where $c$ is a fermion transforming in the fundamental representation. Indeed, the anti-symmetric representation of $\SU(4)$ is six-dimensional so this bi-spinor has the correct number of components. Similarly we can define the dual intertwining matrices $(\overline{\Sigma}^\mu)_{ab}$ and spinors $\tilde{c}^a$ transforming in the anti-fundamental representation. Unlike in the case of $\SU(2)$ the fundamental and anti-fundamental are inequivalent since no tensor can raise or lower indices. The only non-trivial invariant tensor is the four index Levi-Civita tensor $\epsilon_{abcd}$.
	
The Clebsch-Gordan coefficients are defined in terms of the Pauli matrices as 
\begin{equation}
	\begin{split}
		\Sigma^1 &= i\sigma^2 \otimes \sigma^1\,, \qquad\qquad
		\Sigma^2 = -\sigma^1 \otimes \sigma^2\,,\\
		\Sigma^3 &= i\sigma^2 \otimes \sigma^{3}\,,  \qquad\qquad
		\Sigma^4 = \sigma^2 \otimes \mathbf{1}\,,\\
		\Sigma^5 &=  \sigma^3 \otimes \sigma^2\,, \qquad\qquad
		\Sigma^6 = i\sigma^0 \otimes \sigma^2\,.
\end{split}
\end{equation}
These matrices satisfy the following relations
\begin{equation}
	\begin{split}
		\Tr \Sigma_\mu\overline{\Sigma}_\nu &= -4\eta_{\mu\nu}\,,\\
		\left(\Sigma_\mu\right)_{ab}\left(\Sigma^\mu\right)_{cd}	&=2\varepsilon_{abcd}\,,\\
		\left(\overline{\Sigma}_\mu\right)^{ab}\left(\overline{\Sigma}^\mu\right)^{cd}	&=2\varepsilon^{abcd}\,,\\
		\left(\Sigma_\mu\right)_{ab}\left(\overline{\Sigma}^\mu\right)^{cd}	&=2\left(\delta_a^c\delta_b^d-\delta_a^d\delta_b^c\right)\,,
\end{split}
\end{equation}
where $\overline{\Sigma}$ is the complex conjugate of $\Sigma$.
	
\section{The 4d $\cN=2$ superconformal algebra}
\label{app:42}

In this appendix, we collect some basic facts about the 4d $\cN =2$ superconformal algebra.\footnote{See \cite{Frappat:1996pb} for a comprehensive review on superalgebras.} The spacetime symmetry algebra for $\cN=2$ superconformal field theory is the superalgebra $\sl(4|2)$.\footnote{Here and in the next appendix we specify the complexified superalgebra. One should specify a real form together with the correct representation of the $\sigma^\mu$ matrices to obtain the desired real superalgebra and commutation relations.} The maximal bosonic subalgebra is $\so(6,\mathbf{C}) \times \gl(2,\mathbf{C})$. 
	
The four-dimensional complexified conformal algebra $\so(6,\mathbf{C})$ is generated by translations, special conformal transformations, rotations, and dilatations. The generators for these transformations are given by
\begin{equation}
\cP_{\al\ald}\,,\qquad \cK^{\ald\al}\,,\qquad {\cM_\al}^\be\,,\qquad {\cM^\ald}_\bed\,,\qquad \cH\,,
\end{equation}
where we use the bi-spinor notation (see Appendix \ref{App:bispin}). By adding eight Poincar\'e supercharges $\cQ_\al^\cI$, $\widetilde{\cQ}_{\cI\ald}$ and eight conformal supercharges $\cS_I^\al$, $\widetilde{\cS}^{\cI\ald}$ to this algebra we obtain the 4d $\cN=2$ superalgebra $\sl(4|2)$. These supercharges are acted upon by the  $\gl(2,\mathbf{C})\simeq \sl(2,\mathbf{C})_R\times \gl(1,\mathbf{C})_r$ R-symmetry with generators ${\cR^\cI}_\cJ$ where $\cI,\cJ = 1,2$ are indices in the fundamental of $\gl(2,\mathbf{C})$.
	
The commutation relations for the $\so(6,\mathbf{C})$ conformal algebra are
\begin{equation}
	\begin{aligned}
	\left[ {{\cM}_{\alpha}}^{\beta} , {{\cM}_{\gamma}}^{\delta} \right] &= \delta_{\gamma}^{\beta} {{\cM}_{\alpha}}^{\delta}-\delta_\alpha^\delta {{\cM}_{\gamma}}^\beta \,,&
	\left[{\cM^{\dot{\alpha}}}_{\dot{\beta}} , {\cM^{\dot{\gamma}}}_{\dot{\delta}}\right] &= \delta^{\dot{\alpha}}_{\dot{\delta}} {\cM^{\dot{\gamma}}}_{\dot{\beta}}-\delta^{\dot{\gamma}}_{\dot{\beta}} {\cM^{\dot{\alpha}}}_{\dot{\delta}} \,,\\
	\left[ {{\cM}_{\alpha}}^{\beta} , {{\cP}_{\gamma\dot{\gamma}}} \right] &= \delta_\gamma^\beta {\cP}_{\alpha\dot{\gamma}} - \f12 \delta_\alpha^\beta {\cP}_{\gamma\dot{\gamma}}\,,&
	\left[{\cM^{\dot{\alpha}}}_{\dot{\beta}} , {{\cP}_{\gamma\dot{\gamma}}} \right] &= \delta^{\dot{\alpha}}_{\dot{\gamma}}\cP_{\gamma\dot{\beta}}-\f12 \delta^{\dot{\alpha}}_{\dot{\beta}} {\cP}_{\gamma\dot{\gamma}}\,,\\
	\left[ {{\cM}_{\alpha}}^{\beta} , {{\cK}^{\dot{\gamma}\gamma}} \right] &= -\delta_\alpha^\gamma {\cK}^{\dot{\gamma}\beta} + \f12 \delta_\alpha^\beta {\cK}^{\dot{\gamma}\gamma}\,,&
	\left[{\cM^{\dot{\alpha}}}_{\dot{\beta}} , {{\cK}^{\dot{\gamma}\gamma}} \right] &= -\delta^{\dot{\gamma}}_{\dot{\beta}}\cK^{\dot{\alpha}\gamma}+\f12 \delta^{\dot{\alpha}}_{\dot{\beta}} {\cK}^{\dot{\gamma}\gamma}\,,\\
	\left[ \cH , \cP_{\alpha\dot{\alpha}} \right] &= \cP_{\alpha\dot{\alpha}}\,,&
	\left[ \cH , \cK^{\dot{\alpha}\alpha} \right] &= -\cK^{\dot{\alpha}\alpha}\,,\\
	\left[ \cK^{\dot{\alpha}\alpha} , \cP_{\beta\dot{\beta}} \right] &= \delta_\beta^\alpha \delta^{\dot{\alpha}}_{\dot{\beta}}\cH + \delta_\beta^\alpha {\cM^{\dot{\alpha}}}_{\dot{\beta}} + \delta^{\dot{\alpha}}_{\dot{\beta}} {{\cM}_{\beta}}^{\alpha} \,.&&
	\end{aligned}
\end{equation}
The R-symmetry generators are defined as 
\begin{equation}\label{eq:rR4ddef}
{\cR^1}_{2} = \cR^+\,,\qquad {\cR^2}_{1} = \cR^-\,, \qquad {\cR^1}_{1} =  r+R \,,\qquad {\cR^2}_{2} = r - R\,, 
\end{equation}
where $\cR^\pm$ and $R$ form a Chevalley basis of generators for $\sl(2,\mathbf{C})_R$. The commutation relations obeyed by the R-charges are
\begin{equation}
\left[ {\cR^\cI}_\cJ ,{\cR^\cK}_\cL \right] = \delta^\cK_\cJ \cR^\cI_\cL - \delta^\cI_\cL \cR^\cK_\cJ\,.
\end{equation}
The non-vanishing commutators between the supercharges are
\begin{equation}
	\begin{aligned}
	\left\{ \cQ_\al^\cI,\widetilde{\cQ}_{\cJ\ald} \right\} &= \delta^\cI_\cJ \cP_{\al\ald}\,,\\
	\left\{ \widetilde{\cS}^{\cI\ald} , \cS_\cJ^\al\right\} &= \delta^\cI_\cJ \cK^{\ald\al}\,,\\
	\left\{ \cQ_\al^\cI,\cS_{\cJ}^\be \right\} &= \f12\delta^\cI_\cJ\delta_\al^\be\cH + \delta^\cI_\cJ {\cM_\al}^\be - \delta_\al^\be {\cR^\cI}_\cJ\,,\\
	\left\{ \widetilde{\cS}^{\cI\ald},\widetilde{\cQ}_{\cJ\bed} \right\} &= \f12\delta^\cI_\cJ\delta^\ald_\bed\cH + \delta^\cI_\cJ {\cM^\ald}_\bed + \delta^\ald_\bed {\cR^\cI}_\cJ\,,
	\end{aligned}
\end{equation}
Finally, the bosonic generators act on the supercharges as
\begin{equation}
\begin{aligned}
	\left[{\cM_{\al}}^{\be} , \cQ_{\gamma}^{\cI} \right] &= \delta_{\gamma}^{\beta} \cQ_{\alpha}^{\cI} - \f12 \delta_{\al}^{\be} \cQ_{\gamma}^{\cI}\,, & 
	\left[ {\cM^{\ald}}_{\bed} , \widetilde{\cQ}_{\cI\dot{\gamma}} \right] &= \delta_{\dot{\gamma}}^{\ald} \widetilde{\cQ}_{\cI\bed} - \f12 \delta^{\ald}_{\bed}  \widetilde{\cQ}_{\cI\dot{\gamma}}\,,\\
	\left[{\cM_{\al}}^{\be} , \cS_{\cI}^{\gamma} \right] &= -\delta_{\alpha}^{\gamma} \cS_{\cI}^{\beta} + \f12 \delta_{\al}^{\be} \cS_{\cI}^{\gamma}\,, &
	\left[ {\cM^{\ald}}_{\bed} , \widetilde{\cS}^{\cI\dot{\gamma}} \right] &= -\delta^{\dot{\gamma}}_{\bed} \widetilde{\cS}^{\cI\ald} + \f12 \delta^{\ald}_{\bed}  \widetilde{\cS}^{\cI\dot{\gamma}}\,,\\
	\left[ \cH , \cQ_\al^\cI \right] &= \f12 \cQ_\al^\cI \,,&
	\left[ \cH , \widetilde{\cQ}_{\cI\ald} \right] &= \f12 \widetilde{\cQ}_{\cI\ald} \,.\\
	\left[ \cH , \cS_\cI^\alpha \right] &= -\f12 \cS_\cI^\al \,,&
	\left[ \cH , \widetilde{\cS}^{\cI\ald} \right] &= -\f12 \widetilde{\cS}^{\cI\ald} \,.\\
	\left[{\cK^{\ald\al}} , \cQ_{\beta}^{\cI} \right] &= \delta_{\beta}^{\alpha}\widetilde{\cS}^{\cI\ald}\,,&
	\left[ {\cK^{\ald\al}} , \widetilde{\cQ}_{\cI\dot{\beta}} \right] &= \delta_{\dot{\beta}}^{\ald} \cS_{\cI}^\al\,,\\
	\left[{\cP_{\al\ald}} , \cS_{\cI}^{\beta} \right] &= -\delta_{\alpha}^{\beta} \widetilde{\cQ}_{\cI\ald}\,,&
	\left[ {\cP_{\al\ald}} , \widetilde{\cS}^{\cI\dot{\beta}} \right] &= -\delta^{\dot{\beta}}_{\ald} \cQ^{\cI}_\al\,,\\
	\left[ {\cR^\cI}_\cJ , \cQ_\al^\cK \right] &= \delta_\cJ^\cK \cQ_\al^\cI - \f14 \delta_\cJ^\cI \cQ_\al^\cK\,,&
	\left[ {\cR^\cI}_\cJ , \widetilde{\cQ}_{\cK\ald} \right] &= -\delta_\cK^\cI \widetilde{\cQ}_{\cJ\ald} + \f14 \delta_\cJ^\cI \widetilde{\cQ}_{\cK\ald}\,.
	\end{aligned}
\end{equation}	
All other commutators vanish.
	
In radial quantization, the various generators satisfy the following hermiticity conditions
\begin{equation}
	\begin{aligned}
	\cH^\dagger = \cH\,,& \qquad (\cP_{\al\ald})^\dagger = \cK^{\ald\al}\,,\qquad ({{\cM}_\alpha}^\beta)^\dagger = {{\cM}_\beta}^\alpha\,,\qquad ({{\cM}^{\ald}}_{\bed})^\dagger = {{\cM}^{\bed}}_{\ald}\,,\\
	&({{\cR}^{\cI}}_{\cJ})^\dagger = {{\cR}^{\cJ}}_{\cI}\,,\qquad ({\cQ_{\al}^{\cI}})^\dagger = {\cS^{\al}_{\cI}}\,,\qquad ({{\widetilde{\cQ}}_{\cI\ald}})^\dagger = {{\widetilde{\cS}}^{\cI\ald}}\,.
	\end{aligned}
\end{equation}
%
	
\section{The 6d $\cN= (2,0)$ superconformal algebra}
\label{app:20}

In this appendix, we collect basic facts about the 6d $\cN=(2,0)$ superconformal algebra and establish our conventions. The spacetime symmetry algebra is the superalgebra $D(4,2)=\osp(8|4)$. The maximal bosonic subalgebra is $\so(8,\mathbf{C})\times \sp(4,\mathbf{C})$.
	
The six-dimensional complexified conformal algebra $\so(8,\mathbf{C})$ is generated by translations, special conformal transformations, rotations, and dilatations, with respective generators
\begin{equation}
\cP_{ab}\,,\qquad \cK^{ab}\,,\qquad {\cM_a}^b\,,\qquad \cH\,.
\end{equation}
Adding to this 16 Poincar\'e and 16 conformal supercharges, $\cQ_{Aa}$ and $\cS_A^a$, and the generators ${\cR}_{AB}$ of the R-symmetry group $\sp(4,\mathbf{C})$ we obtain the $\cN = (2,0)$ superconformal algebra $\osp(8|4)$. We again use the bi-spinor notation summarized in Appendix \ref{App:bispin} for the indices $a,b=1,\ldots,4$. The $A$ index transforms in the fundamental representation of $\sp(4)$. The $A,B$ indices are raised and lowered with $\Omega_{AB}$, the skew-symmetric symplectic matrix with $\Omega_{14}=\Omega_{23}=1$ and the other entries vanishing. 

We use an oscillator representation of the superconformal algebra \cite{Gunaydin:1985tc}. In addition to the fermionic oscillators $c_a$ and $\tilde{c}^a$ from the bi-spinor formalism we introduce another set of bosonic oscillators $\alpha_A$. These oscillators satisfy the following commutation relations
\begin{equation}
		\{ c_a , \tilde{c}^b \} = \delta_a^b \,,\qquad [\alpha_A,\alpha_B] = \Omega_{AB}\,.
\end{equation}
We can define the generators of the bosonic $\so(8,\mathbf{C})$ algebra as bi-spinors
\begin{equation}	
	\begin{split}
		\cP_{ab} &= c_a c_b\,, \qquad\qquad \cK^{ab} = \tilde{c}^a \tilde{c}^b\,, \\
		{\cM_{a}}^b &= c_a \tilde{c}^b - \f14 \delta_a^b c_d \tilde{c}^d \,,\quad\qquad \cH = \f12 c_a \tilde{c}^a \,.
	\end{split}
\end{equation}	
In particular, note that ${\cM_a}^b$ is traceless. In addition to these bosonic generators, we can define the fermionic and the additional bosonic R-symmetry generators as
\begin{equation}
	\cQ_{Aa} = c_a \alpha_A\,,\qquad S_A^a = \tilde{c}^a\alpha_A \,,\qquad \cR_{AB} = \alpha_A\alpha_B\,.
\end{equation}
The fermionic anti-commutators are given by
\begin{equation}
	\begin{split}
	\{ \cQ_{Aa} , \cQ_{Bb}\} &= \Omega_{AB}\cP_{ab}\,,\\
	\{ \cS_{A}^a , \cS_{B}^b\} &= \Omega_{AB}\cK^{ab}\,,\\
	\{ \cQ_{Aa} , \cS_{B}^b\} &= {\delta_a}^b \cR_{AB} + \Omega_{AB}{\cM_{a}}^b + \f12{\delta_a}^b \Omega_{AB} \cH \,.
	\end{split}
\end{equation}	
The non-vanishing commutators of the bosonic generators read
\begin{equation}
	\begin{split}
	[\cP_{ab},\cK^{cd}] &=\left(\delta_a^c\delta_b^d-\delta_b^c\delta_a^d\right)\cH+ \delta_b^c{\cM_a}^d+\delta_a^d{\cM_b}^c-\delta_a^c{\cM_b}^d-\delta_b^d{\cM_a}^c\,,\\
	[\cP_{ab},{\cM_c}^d] &= \delta_a^d \cP_{bc} -\delta_b^d \cP_{ac} +\f12 \delta_c^d \cP_{ab}\,,\\
	[\cK^{ab},{\cM_c}^d] &= \delta_c^b \cK^{ad} -\delta_c^a \cK^{bd} -\f12 \delta_c^d \cK^{ab}\,,\\
	[{\cM_a}^b,{\cM_c}^d] &= -\delta_a^d {\cM_c}^b + \delta_c^b {\cM_a}^d\,,\\
	[\cH, \cP_{ab}] &= \cP_{ab}\,,\\
	[\cH, \cK^{ab}] &= -\cK^{ab}\,,\\
	[\cR_{AB}, \cR_{CD}] &= \Omega_{AC}\cR_{BD}+\Omega_{BC}\cR_{AD}+\Omega_{AD}\cR_{BC}+\Omega_{BD}\cR_{AC}\,.
	\end{split}
\end{equation}	
Finally, the fermionic and bosonic generators have the following commutation relations
\begin{equation}
	\begin{aligned}
	[\cP_{ab},\cQ_{Cc}] &= 0\,, & [\cP_{ab},\cS_{C}^c] &= \delta_b^c \cQ_{Ca}-\delta_a^c \cQ_{Cb}\,,\\
	[\cK^{ab},\cQ_{Cc}] &= \delta_c^b \cS_{C}^a-\delta_c^a \cS_{C}^b\,, & [\cK^{ab},\cS_{C}^c] &= 0\,,\\
	[{\cM_a}^b,\cQ_{Cc}] &= \delta_c^b\cQ_{Ca} - \f14 \delta_a^b \cQ_{Cc}\,, & [{\cM_a}^b,\cS_{C}^c] &= -\delta_a^c\cS_{C}^b + \f14 \delta_a^b \cS_{C}^c\,,\\
	[\cH , \cQ_{Cc}] &= \f12 \cQ_{Cc}\,, & [\cH , \cS_{C}^c] &= -\f12 \cS_{C}^c\,, \\
	[\cR_{AB} , \cQ_{Cc}] &= \Omega_{AC}\cQ_{Bc} + \Omega_{BC}\cQ_{Ac}\,, &  [\cR_{AB} , \cS_{C}^c] &= \Omega_{AC}\cS_{B}^c + \Omega_{BC}\cS_{A}^c\,.	\end{aligned}
\end{equation}	
In radial quantization, the generators satisfy the following hermiticity conditions
\begin{equation}
\begin{split}
	\cH^\dagger &= \cH\,, \qquad (\cP_{ab})^\dagger = \cK^{ab}\,, \qquad ({{\cM}_a}^b)^\dagger = {{\cM}_b}^a\,,\\
	({\cR}_{AB})^\dagger &= \Omega_{AC}\Omega_{DB}\cR_{DC}\,, \qquad ({\cQ_{Aa}})^\dagger = \Omega_{AB}{\cS_{A}^{a}}\,.
\end{split}
\end{equation}
%

\subsection*{4d $\cN=2$ subalgebra}\label{App:subalg}
	
There is a 4d $\cN=2$ subalgebra of the 6d $\cN=(2,0)$ algebra described above. This subalgebra plays an important role in the construction outlined in Section~\ref{sec:6dchiral}.

The 4d $\cN=2$ Poincar\'e charges are given by the following list
	\begin{align}
	\cQ^1_+ &= \cQ_{31} \,, & \cQ^1_- &= \cQ_{32} \,, & \cQ^2_+ &= \cQ_{41} \,, & \cQ^2_- &= \cQ_{42} \,, \\
	\widetilde{\cQ}_{1\dot{+}} &= -\cQ_{23} \,, & \widetilde{\cQ}_{1\dot{-}} &= -\cQ_{24} \,, & \widetilde{\cQ}_{2\dot{+}} &= -\cQ_{13} \,, & \widetilde{\cQ}_{2\dot{-}} &= -\cQ_{14} \,.
	\end{align}
Similarly, the 4d $\cN=2$ superconformal charges are given by
	\begin{align}
	\cS^+_1 &= -\cS_{2}^1 \,, & \cS_1^- &= -\cS_{2}^2 \,, & \cS_2^+ &= -\cS_{1}^1 \,, & \cS_2^- &= -\cS_{1}^2 \,, \\
	\widetilde{\cS}^{1\dot{+}} &= -\cS_{3}^3 \,, & \widetilde{\cS}^{1\dot{-}} &= -\cS_{3}^4 \,, & \widetilde{\cS}^{2\dot{+}} &= -\cS_{4}^3 \,, & \widetilde{\cS}^{2\dot{-}} &= -\cS_{4}^4 \,.
	\end{align}
The 4d translations and special conformal transformations can be identified as
\begin{align}
	\cP_{+\dot{+}} &= \cP_{13}\,,& \cP_{+\dot{-}} &= \cP_{14}\,,& \cP_{-\dot{+}} &= \cP_{23}\,,& \cP_{-\dot{-}} &= \cP_{24}\,,\\
	\cK^{\dot{+}+} &= \cK^{13}\,,& \cK^{\dot{+}-} &= \cK^{23}\,,& \cK^{\dot{-}+} &= \cK^{14}\,,& \cK^{\dot{-}-} &= \cK^{24}\,.
\end{align}
If we choose a four-dimensional local frame such that the generators of the rotations in the two orthogonal planes $\cL_2$ and $\cL_3$ are given by $({{\cM}_{+}}^{+}-{{\cM}_{-}}^{-})\pm ({{\cM}^{\dot{+}}}_{\dot{+}}-{{\cM}^{\dot{-}}}_{\dot{-}})$, respectively, we can identify
\begin{align}
	{{\cM}_{+}}^{+} &= {\cM_1}^1\,,& {{\cM}_{+}}^{-} &= {\cM_1}^2\,,&  {{\cM}_{-}}^{+} &= {\cM_2}^1\,,&  {{\cM}_{-}}^{-} &= {\cM_2}^2\,,\\
	{{\cM}^{\dot{+}}}_{\dot{+}} &= {\cM_3}^3\,,& {{\cM}^{\dot{+}}}_{\dot{-}} &= {\cM_4}^3\,,&  {{\cM}^{\dot{-}}}_{\dot{+}} &= {\cM_3}^4\,,& {{\cM}^{\dot{-}}}_{\dot{-}} &= {\cM_4}^4\,.
\end{align}
Finally, the R-symmetry generators  read
\begin{equation}
	{\cR^1}_1 = \cR_{23}\,, \qquad {\cR^1}_2 = \cR_{13}\,, \qquad {\cR^2}_1 = \cR_{24} \,, \qquad {\cR^2}_2 = \cR_{14}\,.
\end{equation}
The generators of the $\gl(1,\mathbf{C})_r$-symmetry and the Cartan of the $\sl(2,\mathbf{C})_R$ are given by
\begin{equation}
r =\f12\left({\cR^1}_{1}+{\cR^2}_{2}\right)\,, \qquad R = \f12\left({\cR^1}_{1}-{\cR^2}_{2}\right) \,.
\end{equation}
%

\section{$\Omega$-deformation of the Kapustin twist}
\label{App:OhYagi}

In the construction of \cite{Oh:2019bgz,Jeong:2019pzg} the chiral algebra in \cite{Beem:2013sza} is derived by employing the topological-holomorphic twist of Kapustin \cite{Kapustin:2006hi} on $\Sigma\times C$ such that the theory is topological on $\Sigma$ and holomorphic on $C$. To obtain a non-trivial chiral algebra it is important to perform also an $\Omega$-deformation on $\Sigma$. In \cite{Oh:2019bgz,Jeong:2019pzg} it was assumed that $C$ is simply $\mathbf{R}^2$. As discussed in Section~\ref{subsec:Omega6d} there is a subtlety in constructing a proper $Q_\Omega$ supercharge in this case. There is a simple remedy to this problem which can be applied as in Section~\ref{subsec:Omega6d}. Here we briefly outline how to do this.

In terms of the 4d $\mathcal{N}=2$ supercurrents we can define the $\Omega$-deformed twisted supercharge as $Q_\Omega = Q+ \iota_V G$, where $Q$ is the scalar supercharge of the four-dimensional topological-holomorphic twist of \cite{Kapustin:2006hi} and $G$ is a one-form supercharge associated to the same twist. The subtlety arises due to the fact that $G$ is a one-form on $\Sigma$. This will not be a problem if $C$ is a compact Riemann surface but for $C=\mathbf{R}^2$ one has to be more careful. As in Section~\ref{subsec:Omega6d}, the resolution amounts to contracting $G$ with the four-dimensional Killing vector $V=\Omega_{\mu\nu}x^\nu \partial^\mu$ and thus effectively performing an $\Omega$-deformation on $C$. We find the following result\footnote{The same result can be obtained by following the procedure discussed around Equation (3.33) in \cite{Oh:2019bgz}. We are grateful to Junya Yagi for a useful discussion on this.} 
\begin{equation}\label{eq:vecG4dN2}
\begin{split}
	\iota_V G = i\int \dd^3 x &\left( (\bar{\sigma}_m)^{\dot{\alpha}\beta} \mathcal{G}^\cI_{\beta} M_{\cI \dot{\alpha}}\Omega^{mn}x_n + ({\sigma}_m)^{\alpha \dot{\beta}} \bar{\mathcal{G}}_{\cI\dot{\beta}} \tilde{M}^\cI_{\alpha}\Omega^{mn}x_n\right)\\
	&+\left( (\bar{\sigma}_p^{\dot{\alpha}\beta} \mathcal{G}^\cI_{\beta} N_{\cI \dot{\alpha}}\Omega^{pq}x_q + ({\sigma}_p)^{\alpha \dot{\beta}} \bar{\mathcal{G}}_{\cI\dot{\beta}} \tilde{N}^\cI_{\alpha}\Omega^{pq}x_q\right) = 
	\epsilon(\widetilde{\cS}^{2\dot{-}}-{\cS_1}^{-})
\end{split}
\end{equation}
where $m,n=1,2$ and $p,q=3,4$ and the matrices encoding the topological-holomorphic twist are given by
\begin{equation}
	M=	
	\begin{pmatrix}
		0&1\\0&0
	\end{pmatrix}
	\,,\quad 
	\tilde{M}=	
	\begin{pmatrix}
		0&0\\1&0
	\end{pmatrix}\,, \quad
	N=	
	\begin{pmatrix}
		0&0\\0&-1
	\end{pmatrix}
	\,,\quad 
	\tilde{N}=	
	\begin{pmatrix}
		0&0\\1&0
	\end{pmatrix}\,.
\end{equation}
The matrix $\Omega$ is given by 
\begin{equation}
\Omega = \begin{pmatrix}
			0&\epsilon&0&0\\-\epsilon&0&0&0\\0&0&0&\epsilon\\0&0&-\epsilon&0
		\end{pmatrix}\,.
\end{equation}
This is simply the familiar $\Omega$-deformation in four dimensions with $\epsilon_1=\epsilon_2$. 

Using that the scalar supercharge of the topological-holomorphic twist in \cite{Kapustin:2006hi} is given by $Q=\mathcal{Q}^1_- + \widetilde{\mathcal{Q}}_{2\dot{-}}$ together with the result in \eqref{eq:vecG4dN2} we find the following $\Omega$-deformation supercharge
\begin{equation}
Q_{\Omega} =\mathcal{Q}^1_- + \widetilde{\mathcal{Q}}_{2\dot{-}} +\epsilon (\widetilde{\mathcal{S}}^{2\dot{-}} - \mathcal{S}_1^-)\,.
\end{equation}
This is precisely the supercharge used in \cite{Oh:2019bgz,Jeong:2019pzg} where it was shown that passing to its cohomology leads to the chiral algebra derived in \cite{Beem:2013sza}.

\section{2d $\mathcal{N}=(0,4)$ SCFTs and chiral algebras}
\label{App:N4SCA}
	
The chiral algebra construction of \cite{Beem:2013sza} can be applied also to 2d SCFTs with $\mathcal{N}=(0,4)$ supersymmetry. The goal of this appendix is to review the small $\mathcal{N}=(0,4)$ superconformal algebra and to show that the operators spanning the chiral algebra resulting from the construction in \cite{Beem:2013sza} are simply those in the left-moving, i.e. non-supersymmetric, sector of the $\mathcal{N}=(0,4)$ SCFT.

\subsection*{The small $\mathcal{N}=(0,4)$ superconformal algebra}	
	
Here we collect some well-known facts about the small $\mathcal{N}=4$ superconformal algebra in two dimensions. The global part of the spacetime symmetry algebra is given by the superalgebra $\sl(2)_l\times \su(1,1|2)$ which has the maximal bosonic subalgebra $\sl(2)_l\times\sl(2)_r\times \su(2)_R$. In this appendix, we work to Euclidean signature.
	
On the Euclidean plane, with complex coordinate $z$, the stress-energy tensor has two independent components, $T(z)$ and $\bar{T}(\bar{z})$ with the following Laurent series representation
\begin{equation}
T(z) = \sum_{n=-\infty}^{\infty} \f{L_n}{z^{n+2}}\,, \qquad \bar{T}(\bar{z}) = \sum_{n=-\infty}^{\infty} \f{\bar{L}_n}{\bar{z}^{n+2}}\,.
\end{equation}
The Laurent coefficients $L_n$ and $\bar{L}_n$ generate two copies of the Virasoro algebra with central charge $c$.\footnote{In general the left- and right-moving Virasoro algebras may have different central charges.} In addition to the energy-momentum tensor, the small $\cN=4$ algebra contains four spin-$\f32$ currents $\bar{G}^{aA}$, where $a$ is a doublet index with respect to the $\su(2)_R$ R-symmetry current algebra. The integer level, $k$, of the current algebra is related to the central charge $c=6k$. The index $A$ is a doublet under an outer automorphism $\SU(2)_{\rm out}$ which is not part of the superconformal algebra and in general not a symmetry of the SCFT. The supercurrents admit the Laurent series expansion
\begin{equation}
\bar{G}^{aA}(\bar{z}) = \sum_{r\in \mathbf{Z}+\f12} \f{\bar{G}^{aA}_r}{\bar{z}^{r+\f32}}\,.
\end{equation}
Finally, the $\su(2)_R$ R-symmetry is generated by the currents $\bar{J}^i$ which have the Laurent series expansion
\begin{equation}
\bar{J}^i(\bar{z}) = \sum_{p=-\infty}^{\infty} \f{\bar{T}^i_p}{\bar{z}^{p+1}}\,.
\end{equation}

The non-trivial (anti-)commutation relations among the right-moving modes close the small $\mathcal{N}=4$ superconformal algebra\footnote{See for example \cite{Schwimmer:1986mf}. However, note that there are some typos in that reference.}
\begin{equation}\label{N=4SCA}
	\begin{split}
	[\bar{L}_m, \bar{L}_n ] &=  (m-n) \, \bar{L}_{m+n} + \tfrac{1}{12}c(m^3-m)\delta_{m+n,0}, \\
	[\bar{T}^i_m,\bar{T}^j_n] &= i \epsilon_{ijk} \bar{T}^k_{m+n}+k\,m\,\delta_{m+n,0}\delta^{i,j} , \\
	[\bar{L}_m,\bar{T}^i_n] &= -n \bar{T}^i_{m+n}, \\
	[\bar{L}_m, \bar{G}_r^{aA}] &= \left( \tfrac m2 - r \right) \bar{G}^{aA}_{m+r}, \\ 
	[\bar{T}_i, \bar{G}_r^{a1}] &= \tfrac{1}{2}\sigma_{ba}^{i} \bar{G}^{b1}_{r}, \\
	[\bar{T}_i, \bar{G}_r^{a2}] &= -\tfrac{1}{2} \overline{\sigma}_{ba}^{i}\bar{G}^{b2}_{r}, \\
	\{ \bar{G}_r^{a1}, \bar{G}_s^{b1}\} &= \{ \bar{G}_r^{a2}, \bar{G}_s^{b2} \} = 0,\\
	\{ \bar{G}_r^{a1}, \bar{G}_s^{b2} \} &= 2 \delta_{ab}\bar{L}_{r+s} + 2 \sigma_{ab}^{i}(r-s) \bar{T}^i_{r+s}+\tfrac{1}{3}c(r^2-\tfrac{1}{4})\delta_{r+s,0}\delta_{ab}\,,
	\end{split}
	\end{equation}	
where $\sigma^i$ and $\bar{\sigma}^i$ are the Pauli matrices and their complex conjugates \eqref{eq:Paulisigma}. Similarly, the Laurent coefficients of the holomorphic component of the energy-momentum tensor, $L_{m}$ generate the left-moving Virasoro algebra with commutation relations 
\begin{equation}
[L_m, L_n ] =  (m-n) \, L_{m+n} + \tfrac{1}{12}c(m^3-m)\delta_{m+n,0}\,,
\end{equation} 
and commute with all the right-moving generators. 
	
We are interested in the algebra $\mathfrak{su}(1,1|2)$ which corresponds to the right-moving part of the global superconformal algebra above. The global $\sl(2)_l\times \sl(2)_r$ algebra is generated by $L_0$, $L_{\pm1}$ and $\bar{L}_0$, $\bar{L}_{\pm1}$ and $\su(2)_R$ is generated by $\bar{T}_0^i$. The global Poincar\'e supercharges are given by $\bar{G}^{a1}_{-\f12}$ and $\bar{G}^{b2}_{-\f12}$ and the conformal supercharges by $\bar{G}^{a1}_{\f12}$ and $\bar{G}^{b2}_{\f12}$. It is useful to rename the global supercharges and R-symmetry generators as as
	\begin{align}
	\mathcal{Q}^1 &\equiv \tfrac{1}{\sqrt{2}} \bar{G}_{-\frac{1}{2}}^{11}\,, & \mathcal{Q}^2 &\equiv \tfrac{1}{\sqrt{2}} \bar{G}_{-\frac{1}{2}}^{21}\,, & \widetilde{\mathcal{Q}}_1 &\equiv \tfrac{1}{\sqrt{2}} \bar{G}_{-\frac{1}{2}}^{12}\,, & \widetilde{\mathcal{Q}}_2 &\equiv \tfrac{1}{\sqrt{2}} \bar{G}_{-\frac{1}{2}}^{22}\,, \\
	\mathcal{S}_1 &\equiv \tfrac{1}{\sqrt{2}} \bar{G}_{\frac{1}{2}}^{12}\,, & \mathcal{S}_2 &\equiv \tfrac{1}{\sqrt{2}} \bar{G}_{\frac{1}{2}}^{22}\,, & \widetilde{\mathcal{S}}^1 &\equiv \tfrac{1}{\sqrt{2}} \bar{G}_{\frac{1}{2}}^{11}\,, & \widetilde{\mathcal{S}}^2 &\equiv \tfrac{1}{\sqrt{2}} \bar{G}_{\frac{1}{2}}^{21}\,, \\
	\mathcal{R}^+ &\equiv \bar{T}^1_0+i\bar{T}^2_0\,, & \mathcal{R}^- &\equiv \bar{T}^1_0-i\bar{T}^2_0, & \mathcal{R} &\equiv \bar{T}^3_0\,. & &
	\end{align}
The non-vanishing bosonic commutation relations of $\su(1,1|2)$ are then given by
	\begin{align}
		[\mathcal{R}, \mathcal{R}^\pm] &=  \pm \mathcal{R}^\pm\,, & [\mathcal{R}^+,\mathcal{R}^-] &= 2\mathcal{R}\,, \\
		[\bar{L}_0,\bar{L}_{\pm 1}] &= \mp \bar{L}_{\pm 1}\,, & [\bar{L}_{1}, \bar{L}_{-1}] &= 2\bar{L}_0\,. 
	\end{align}
The anti-commutation relations among the fermionic generators are given by\footnote{The last of these relations differs by a minus sign from the one mentioned in \cite{Beem:2013sza} but matches with \cite{Schwimmer:1986mf} and gives the correct twisted $\widehat{\mathfrak{su}(2)}$.}
	\begin{align}
	\{\mathcal{Q}^{\mathcal{I}}, \widetilde{\mathcal{Q}}_{\mathcal{J}}\} &=  \delta^{\mathcal{I}}_{\mathcal{J}} \bar{L}_{-1}\,,  &
	\{\widetilde{\mathcal{S}}^{\mathcal{I}}, \mathcal{S}_{\mathcal{J}}\} &= \delta^\mathcal{I}_\mathcal{J} \bar{L}_{+1}, \\
	\{\mathcal{Q}^{\mathcal{I}}, \mathcal{S}_{\mathcal{J}}\} &= \delta^\mathcal{I}_\mathcal{J} \bar{L}_{0} - \mathcal{R}^{\mathcal{I}}_\mathcal{J} - \frac{1}{2}\delta^\mathcal{I}_\mathcal{J} \mathcal{Z}\,, &
	\{\widetilde{\mathcal{Q}}_{\mathcal{J}},\widetilde{\mathcal{S}}^{\mathcal{I}}\} &= \delta^{\mathcal{I}}_{\mathcal{J}} \bar{L}_{0} + \mathcal{R}^{\mathcal{I}}_{\mathcal{J}} + \frac{1}{2}\delta^\mathcal{I}_\mathcal{J} \mathcal{Z}\,.
	\end{align}
Where $\mathcal{R}^{\mathcal{I}}_{\mathcal{J}}$ are defined as
	\begin{equation}
	\mathcal{R}^{1}_{2} = \mathcal{R}^+, \qquad \mathcal{R}^{2}_{1} = \mathcal{R}^-, \qquad \mathcal{R}^{1}_{1} = \mathcal{R}, \qquad \mathcal{R}^{2}_{2} = -\mathcal{R}\,. 
	\end{equation}
The generator $\mathcal{Z}$ is a central element which can be added to the algebra but will play no role in our discussion. If we remove $\mathcal{Z}$ we obtain the algebra $\mathfrak{psu}(1,1|2)$. The remaining non-trivial commutators in the algebra are given by 
	\begin{align}
		[\bar{L}_{-1}, \widetilde{\mathcal{S}}^\mathcal{I}] &= - \mathcal{Q}^\mathcal{I}\,, & [\bar{L}_{-1}, \mathcal{S}_\mathcal{I}] &= - \widetilde{\mathcal{Q}}_\mathcal{I}\,, \\
		[\bar{L}_{1}, \widetilde{\mathcal{Q}}_\mathcal{I}] &=  \mathcal{S}_\mathcal{I}\,, & [\bar{L}_{1}, \mathcal{Q}^\mathcal{I}] &=  \widetilde{\mathcal{S}}^\mathcal{I}\,, \\
		[\bar{L}_{0}, \widetilde{\mathcal{S}}^\mathcal{I}] &= -\frac{1}{2} \widetilde{\mathcal{S}}^\mathcal{I}\,, & [\bar{L}_{0}, \mathcal{S}_\mathcal{I}] &= -\frac{1}{2} \mathcal{S}_\mathcal{I}\,, \\
		[\bar{L}_{0}, \widetilde{\mathcal{Q}}_\mathcal{I}] &= \frac{1}{2}\widetilde{\mathcal{Q}}_\mathcal{I}\,, & [\bar{L}_{0}, {\mathcal{Q}}^\mathcal{I}] &= - {\mathcal{Q}}^\mathcal{I}\,.
	\end{align}
	
\subsection*{Chiral algebras from 2d $\cN=(0,4)$ SCFTs} 
\label{subsec:2dtwist}
		
Next, following \cite{Beem:2013sza,Beem:2014kka}, we define a set of special nilpotent supercharges $\mathbbmtt{Q}$ which commute with the generators $L_0,L_{\pm1}$ and for which the anti-holomorphic transformations are $\mathbbmtt{Q}$-exact. The first part of this requirement is trivial, since all right-moving supercharges commutes with the left-moving Virasoro generators. The $\mathbbmtt{Q}$-exactness criterion is more stringent. In fact, up to similarity transformations, the only choices are:
\begin{align}\label{specialQ}
	\mathbbmtt{Q}_1 &= \cQ^1+\widetilde{\cS}^2\,,&\mathbbmtt{Q}_1^\dagger &= \cS_1+\widetilde{\cQ}_2\,,\\
	\mathbbmtt{Q}_2 &= \widetilde{\cQ}_2-\cS_1\,,&\mathbbmtt{Q}_2^\dagger &= \widetilde{\cS}^2-\cQ^1\,,
\end{align}
Both these supercharges give rise to the same $\mathbbmtt{Q}$-exact generators of the anti-holomorphic $\sl(2)$ algebra
\begin{equation}
	\begin{aligned}
		\widehat{L}_{-1} &\equiv \{\mathbbmtt{Q}_1,\widetilde{\mathcal{Q}}_1\} = \{\mathbbmtt{Q}_2,\mathcal{Q}^2\} = \bar{L}_{-1}+ \mathcal{R}^{+}~, \\
		\widehat{L}_{+1} &\equiv \{\mathbbmtt{Q}_1,\mathcal{S}_2\} = \{\mathbbmtt{Q}_2,-\widetilde{\mathcal{S}}^1\} = \bar{L}_{+1}- \mathcal{R}^{-}~, \\
		2\widehat{L}_{0} &\equiv \{\mathbbmtt{Q}_1,\mathbbmtt{Q}_1^{\dagger}\} = \{\mathbbmtt{Q}_2,\mathbbmtt{Q}_2^{\dagger}\} = 2(\bar{L}_{0}- \mathcal{R})~. 
		\end{aligned}
\end{equation}
In addition, we note that the central element of $\mathfrak{su}(1,1|2)$ is exact with respect to both supercharges
	\begin{equation}
		\{\mathbbmtt{Q}_1,\mathbbmtt{Q}_2\}= -\mathcal{Z}\,.
	\end{equation}
With this at hand we can proceed to construct a chiral algebra with operators that are in the cohomology of $\mathbbmtt{Q}_1$ (or equivalently $\mathbbmtt{Q}_2$). Our task thus consists of identifying those operators that define non-trivial cohomology classes. For an operator $\mathcal{O}(x)$ to define a non-trivial $\mathbbmtt{Q}_i$-cohomology class it has to obey the following conditions,
	\begin{equation}\label{cohom}
		\{\mathbbmtt{Q}_i,\mathcal{O}(0)]=0~, \qquad \text{and } \quad \mathcal{O}(0)\neq \{\mathbbmtt{Q}_i,\mathcal{O}^\prime(0)],
	\end{equation}
for $i= 1$ or $2$. Since both $\mathbbmtt{Q}_i$ commute with $\widehat{L}_0$ and $\mathcal{Z}$, we lose no generality by restricting to definite eigenspaces of these charges. A standard cohomological argument then implies that since $\widehat{L}_0$ and $\mathcal{Z}$ are $\mathbbmtt{Q}_i$-exact, an operator satisfying \eqref{cohom} must lie in the zero eigenspace of both charges. Such operators obey 
	\begin{equation}
		\widehat{h}=\bar{h}-r=0~,
	\end{equation} 
where $r$ is the eigenvalue of $\mathcal{R}$. These are precisely the operators contributing to the chiral algebra characterizing the so-called half-twisted model of a $\cN=(0,2)$ theory \cite{Witten:1991zz}.
	
Hence, it appears that we have found a novel chiral algebra inside a $\cN=(0,4)$. However, as already noted above we see that the operators contributing to the cohomology are identical as the ones in the chiral algebra of a half-twisted $\cN=(0,2)$ theory, which is obtained by passing to the cohomology with respect to a Poincar\'e supercharge. Indeed, from the equality
\begin{equation}
	\{\cQ^1,\cS_1 \} = \{ \widetilde{\cQ}_2, \widetilde{\cS}_2 \} = \bar{L}_0-\cR\,.
\end{equation}
we see that going to the cohomology with respect to $\cQ^1$ or $\widetilde{\cQ}_2$ selects the same operators. Hence we have found that the cohomologies are isomorphic as graded vector spaces. However, this isomorphism on the level of graded vector spaces does not necessarily lift to an isomorphism of the full chiral algebra. Indeed, due to the twisted translations the OPE between various operators in the cohomology can still differ from the OPE of the chiral algebra discussed in \cite{Witten:1991zz}.\footnote{We are grateful to Chris Beem for emphasizing this possibility.} 


\bibliography{TheRealOmega}
\bibliographystyle{JHEP}
	
\end{document}